\documentclass{aa}  
\usepackage{amsfonts,amsmath,amssymb}
\usepackage{graphicx}
\usepackage{color}
\usepackage{url}
\usepackage{txfonts}
\usepackage{hyperref}

\newcommand{\p}[1][]{ p\left( #1 \right) } 
\DeclareMathOperator{\given}{|} 
\DeclareMathOperator{\ms}{\,m\,s^{-1}} 
\DeclareMathOperator{\earthmass}{M_{\oplus}} 
\newcommand{\rhk}{$\log R'_{\rm HK}$\xspace}

\begin{document}

   \title{The HARPS search for southern extra-solar planets
          \thanks{Based on observations collected at ESO facilities under programs 082.C-0212, 085.C-0063, 086.C-0284, and 190.C-0027 (with the HARPS spectrograph at the ESO 3.6-m telescope, La Silla-Paranal Observatory).}}

   \subtitle{XL. Searching for Neptunes around metal-poor stars}

   \author{J. P. Faria\inst{1,2}
          \and N. C. Santos\inst{1,2}
          \and P. Figueira\inst{1}
          \and A. Mortier\inst{3}
          \and X. Dumusque\inst{4}
          \and I. Boisse\inst{5}
          \and G. Lo Curto\inst{6}
          \and C. Lovis\inst{7}
          \and M. Mayor\inst{7}
          \and C. Melo\inst{6}
          \and F. Pepe\inst{7}
          \and D. Queloz\inst{7,8}
          \and A. Santerne\inst{1}
          \and D. Ségransan\inst{7}
          \and S. G. Sousa\inst{1}
          \and A. Sozzetti\inst{9}
          \and S. Udry\inst{7}
          }

   \institute{
        Instituto de Astrofísica e Ciências do Espaço, Universidade do Porto, CAUP, Rua das Estrelas, PT4150-762 Porto, Portugal\\
        \email{joao.faria@astro.up.pt}
        \and
        Departamento\,de\,Física\,e\,Astronomia,\,Faculdade\,de\,Ciências,\,Universidade\,do\,Porto,\,Rua\,Campo\,Alegre,\,4169-007\,Porto,\,Portugal
        \and
        SUPA, School of Physics and Astronomy, University of St Andrews, St Andrews KY16 9SS, UK
        \and
        Harvard-Smithsonian Center for Astrophysics, 60 Garden Street, Cambridge, MA 02138, USA
        \and
        Aix Marseille Université, CNRS, LAM (Laboratoire d’Astrophysique de Marseille) UMR 7326, 13388, Marseille, France
        \and
        European Southern Observatory, Casilla 19001, Santiago, Chile
        \and
        Observatoire de Genève, Université de Genève, 51 ch. des Maillettes, CH-1290 Sauverny, Switzerland
        \and
        Institute of Astronomy, University of Cambridge, Madingley Road, Cambridge, CB3 0HA, UK
        \and
        INAF - Osservatorio Astrofisico di Torino, Via Osservatorio 20, I-10025 Pino Torinese, Italy
        }

   \date{Received October 8, 2015; accepted January 7, 2016}


\abstract
  {Stellar metallicity -- as a probe of the metallicity of proto-planetary disks -- is an important ingredient for giant planet formation, likely through its effect on the timescales in which rocky/icy planet cores can form. Giant planets have been found to be more frequent around metal-rich stars, in agreement with predictions based on the core-accretion theory. In the metal-poor regime, however, the frequency of planets, especially low-mass planets, and how it depends on metallicity are still largely unknown.}
  {As part of a planet search programme focused on metal-poor stars, we study the targets from this survey that were observed with HARPS on more than 75 nights. 
  The main goals are to assess the presence of low-mass planets and provide a first estimate of the frequency of Neptunes and super-Earths around metal-poor stars.}
  {We perform a systematic search for planetary companions, both by analysing the periodograms of the radial-velocities and by comparing, in a statistically-meaningful way, models with an increasing number of Keplerians.}
  {A first constraint on the frequency of planets in our metal-poor sample is calculated considering the previous detection (in our sample) of a Neptune-sized planet around HD175607 and one candidate planet (with an orbital period of 68.42\,d and minimum mass $M_p \sin i = 11.14 \pm 2.47 \earthmass$) for HD87838, announced in the present study. This frequency s determined to be close to 13\% and is compared with results for solar-metallicity stars.}
  {}


  \maketitle


\section{Introduction}

More than 600 exoplanets were discovered and many others confirmed using radial-velocity (RV) observations \citep[\url{www.exoplanet.eu},][]{Schneider2011}. 
The method is more sensitive to massive planets orbiting close to their stars since Neptune- and Earth-mass planets induce very small amplitude radial-velocity signals, often at the level of the current observational uncertainties of 1$\ms$ and smaller.
In addition, stellar activity can induce false-positive signals that mimic and mask the radial velocity signature of a low-mass planet \citep[e.g.][]{Bonfils2007,Robertson2014,Santos2014}.
Only with very high precision Doppler spectroscopy and dense sampling one is able to detect these planets \citep[e.g.][]{Dumusque2012,Hatzes2014} and to push the detection limits in the direction of a possibly habitable Earth-like planet, the same stars must be followed for a very long time.

The increasing number of detected planets provides constraints for models of planet formation and evolution \citep[e.g.][]{Udry2007a,Ida2004,Mordasini2009}. 
In particular, the metallicity and, in general, the chemical abundances of the host star, are now known to be a key ingredient in planet formation \citep{Ida2004,Mordasini2015}.
According to the core-accretion theory, proto-planetary disks with higher metallicity are able to form rocky/icy cores in time for runaway accretion to lead to the formation of a giant planet before disk dissipation occurs. 
In lower metallicity disks, the cores do not grow fast enough to accrete gas in large quantities before disk dissipation occurs, implying a lower fraction of giant planets.

\citet{Gonzalez1997} showed the first observational hints for a correlation between the presence of giant planets and metallicity. 
As more and more planets were discovered, this correlation was solidly confirmed: it is more likely to detect a giant planet orbiting a metal-rich star \citep{Santos2004,Fischer2005,Sousa2011a}. 
This result was also confirmed in data from transit surveys \cite[e.g.][]{Buchhave2012}.
Furthermore, it is also now known that the metallicity, or even specific chemical abundance ratios, can have a strong impact on the planet formation efficiency, composition, and architecture \citep{Guillot2006,Adibekyan2013,Adibekyan2015,Dawson2015,Santos2015}.

Because giant planets are more frequent around metal-rich stars, several RV surveys became biased towards metal-rich samples \citep{Tinney2002,Fischer2005,Melo2007,Jenkins2013a}. 
Nevertheless, some programmes focused on the metal-poor regime, and tried to determine the frequency of giant planets orbiting metal-poor stars, and the metallicity limit below which no giant planets can be observed \citep{Sozzetti2009,Santos2011,Mortier2012,Mortier2013}. Until recently, however, these programs focused on the search for massive, giant planets.

Neptune-mass planets, in contrast, have been found to have a relatively flat metallicity distribution \citep{Udry2006,Sousa2008,Sousa2011a,Neves2013}.
In systems with only hot Neptunes, the metallicity distribution actually becomes slightly metal-poor \citep{Mayor2011,Buchhave2012}.
These results are supported by theoretical models which predict lower-mass planets to be common around stars with a wide range of metallicities. 
One should note here the works of \citet{Wang2014} and \citet{Buchhave2015}, who find conflicting results over the existence of a universal planet-metallicity correlation extending to the terrestrial planets, and \citet{Adibekyan2012} who show that, in the metal-poor regime, planets are more prevalent on $\alpha$-enhanced stars (i.e. of higher global metallicity).

The number of detected low-mass planets orbiting metal-poor stars is nevertheless still small.
In an effort to detect these planets and explore the low metallicity regime, our team started a Large Program with HARPS to follow a sample of about 100 metal-poor stars. 
The program and sample were presented in \citet{Santos2014}, to which we point the reader for more details. 
The main goal of the survey is to derive observational constraints on the frequency of Neptunes and super-Earths in the metal-poor regime. 
These estimates will be compared with the ones from the HARPS-GTO program \citep{Mayor2011} which searched for very low mass planets orbiting a sample of solar-neighbourhood stars (thus with solar metallicities). 
When combined, the two surveys will set important constraints for planet formation and evolution models and help in providing an estimate of the frequency of planets in our Galaxy.

This paper presents the analysis of the stars in the metal-poor sample that were observed on more than 75 nights. 
Some of these stars now have precise radial velocity observations covering a baseline of over ten years. 
In Sect \ref{sec:sample} we present the stellar parameters for these stars and their radial-velocity observations. 
The method used to search for Keplerian signals is outlined in Sect. \ref{sec:method} and in Sect \ref{sec:results} the results of its application are presented. Section \ref{sec:comparison_model_selection} presents a simple comparison of the model selection criteria we consider and detection limits for individual stars are derived in Sect. \ref{sec:detection-limits}. 
A first constraint on the occurrence of planets in this sample and our conclusions are presented in Sect. \ref{sec:discussion}.

\section{Sample and Observations}\label{sec:sample}

Our complete sample of metal-poor stars contains 109 targets, chosen from a survey for giant planets orbiting metal-poor stars \citep{Santos2007} and a programme to search for giant planets orbiting a volume-limited sample of FGK dwarfs \citep{Naef2007}. 
The criteria for the definition of this sample are detailed in \citet{Santos2014}.
Stellar parameters were derived for the complete sample from a set of high-resolution HARPS spectra \citep{Sousa2011,Santos2014}.

In this work we analyse those stars which, up to December 2014, were observed on more than 75 nights. 
This number of nights, though not strictly motivated, means that in the worst case scenario we have six data points per parameter, for a model with two planets.
A total of 15 stars meet this criteria. 
In Table \ref{table:parameters} we present the stellar parameters for this subsample.

The activity level of the stars, denoted here by the weighted average of the \rhk values \citep{Noyes1984a}, was derived from the analysis of the CaII H and K lines in the HARPS spectra \citep[e.g.][]{Santos2000,Lovis2011}. 
Estimates for the rotation period of each star were derived from the activity-rotation calibrations of \citet{Noyes1984a} and \citet{Mamajek2008}.
The typical error bar on these estimates, computed from the dispersion in the \rhk time series, is of the order of the difference between the estimates from the two calibrations.
For HD31128, the rotation period derived in this way is around 2.8\,d. 
The very low metallicity of this star, however, places it outside the calibration range of the \rhk flux calibration, and of the adopted activity-rotation relations. 
This renders the rotation period estimate uncertain, and we chose to omit it from the table.

Table \ref{table:timeseries} shows the number of nights each star was observed, the mean error bar $\bar{\sigma_i}$, and the weighted standard deviation of the radial velocities $s_{\text{RV}}$.
The ratio of the two latter quantities gives an indication of whether the radial-velocities vary in excess of the internal errors, and by how much.
This table also shows the baseline of observations.

\begin{figure}
\centering
\includegraphics[width=\hsize]{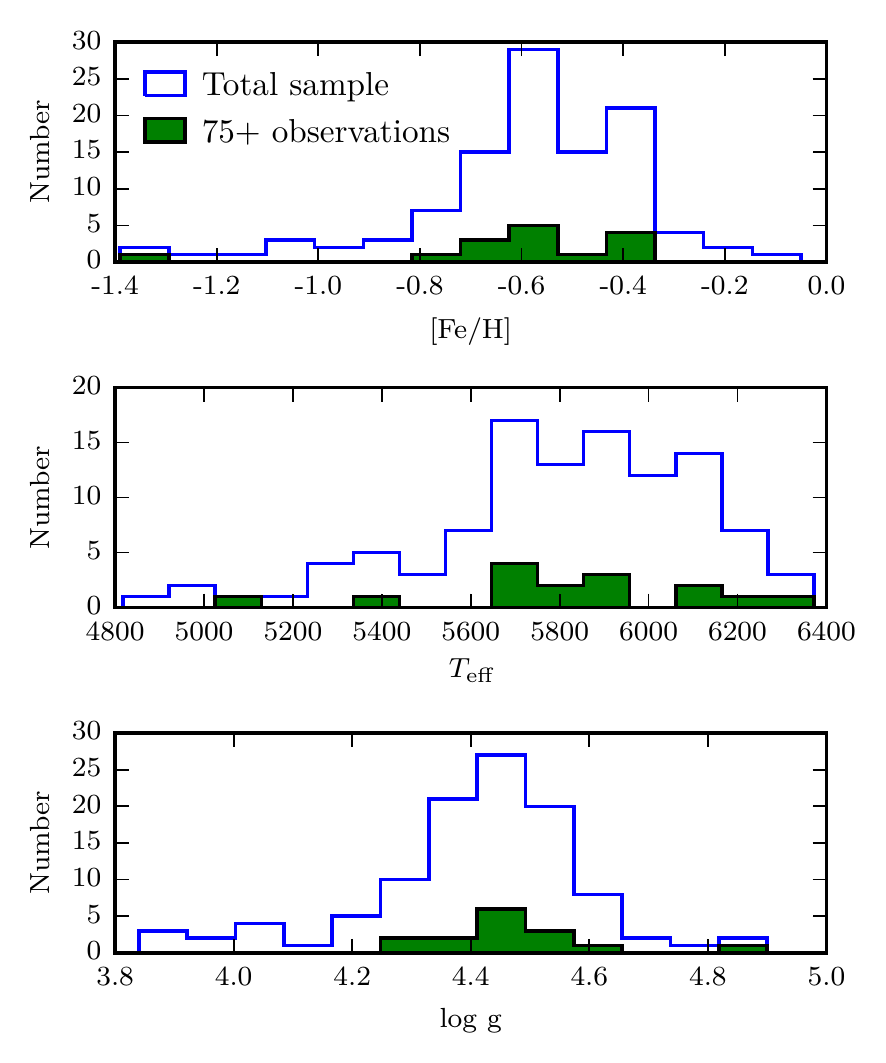}
  \caption{Metallicity, effective temperature and surface gravity distributions for the full metal-poor sample (blue histograms) and the 15 stars observed on more than 75 different nights (green filled histograms).}
     \label{fig:sample-distribution}
\end{figure}

The distributions of metallicity, effective temperature and surface gravity of the full sample and of the 15 stars studied here are shown in Fig. \ref{fig:sample-distribution}.
By selecting stars based on the number of observations, one can introduce biases that depend on the criteria used for scheduling observations.
From the distributions in Fig. \ref{fig:sample-distribution}, we do not see a clear bias towards warmer or more evolved stars, which may show more significant (stellar) variability.
The sub-sample can nevertheless be biased in favour of stars hosting low mass planets, since these stars will also show more significant variability.
This can have an impact on our determination of the planet frequency (Sect. \ref{sec:discussion}), although it is difficult to determine to which extent.
Regarding metallicity, the subsample is representative of the overal distribution but does not span the complete metallicity range.

\begin{table*}
\caption{Stellar parameters for each of the 15 stars studied here. The table lists the mass, effective temperature, surface gravity, metallicity, B-V color, activity level, and estimated rotation period.}
\label{table:parameters}      
\centering          
\begin{tabular}{lcccccccccc}
\hline
\hline

Star & Mass & $T_{\text{eff}}$ & $\log g$ & [Fe/H] & $B-V$ & $\log R'_{HK}$ & $P_{rot}$\tablefootmark{(a)}\\
{\tiny } & {\tiny [$M_\odot$]} & {\tiny [K]} & {\tiny [cgs]} & {\tiny } & {\tiny } & {\tiny } & {\tiny [days]}\\
\hline
HD224817 & 0.98 $\pm$ 0.06 & 5894 $\pm$ 22 & 4.36 $\pm$ 0.02 & -0.53 $\pm$ 0.02 & 0.55 & -4.96 & 13.0 / 13.5 \\
HD21132 & 1.05 $\pm$ 0.07 & 6243 $\pm$ 34 & 4.60 $\pm$ 0.05 & -0.37 $\pm$ 0.02 & 0.49 & -4.90 & 7.2 / 7.3 \\
HD22879 & 0.88 $\pm$ 0.06 & 5884 $\pm$ 33 & 4.52 $\pm$ 0.03 & -0.81 $\pm$ 0.02 & 0.56 & -4.91 & 12.8 / 13.1 \\
HD31128 & 0.79 $\pm$ 0.05 & 6096 $\pm$ 67 & 4.90 $\pm$ 0.05 & -1.39 $\pm$ 0.04 & 0.41 & -4.88 \tablefootmark{(b)} & -- \\
HD41248 & 0.94 $\pm$ 0.06 & 5713 $\pm$ 21 & 4.49 $\pm$ 0.03 & -0.37 $\pm$ 0.01 & 0.61 & -4.89 & 18.2 / 18.4 \\
HD56274 & 0.90 $\pm$ 0.06 & 5734 $\pm$ 22 & 4.51 $\pm$ 0.03 & -0.54 $\pm$ 0.02 & 0.57 & -4.85 & 13.3 / 13.3 \\
HD79601 & 0.94 $\pm$ 0.06 & 5834 $\pm$ 25 & 4.37 $\pm$ 0.04 & -0.60 $\pm$ 0.02 & 0.53 & -4.95 & 11.0 / 11.4 \\
HD87838 & 1.03 $\pm$ 0.07 & 6118 $\pm$ 33 & 4.47 $\pm$ 0.03 & -0.40 $\pm$ 0.02 & 0.52 & -4.92 & 9.7 / 9.9 \\
HD88725 & 0.87 $\pm$ 0.06 & 5654 $\pm$ 17 & 4.49 $\pm$ 0.03 & -0.64 $\pm$ 0.01 & 0.60 & -4.90 & 17.3 / 17.6 \\
HD111777 & 0.87 $\pm$ 0.06 & 5666 $\pm$ 19 & 4.46 $\pm$ 0.03 & -0.68 $\pm$ 0.01 & 0.61 & -4.91 & 18.7 / 19.1 \\
HD114076 & 0.80 $\pm$ 0.06 & 5069 $\pm$ 52 & 4.32 $\pm$ 0.09 & -0.47 $\pm$ 0.04 & 0.82 & -4.98 & 41.4 / 43.4 \\
HD119173 & 0.96 $\pm$ 0.06 & 5779 $\pm$ 44 & 4.26 $\pm$ 0.04 & -0.62 $\pm$ 0.03 & 0.56 & -4.87 & 12.7 / 12.8 \\
HD119949 & 1.09 $\pm$ 0.07 & 6359 $\pm$ 36 & 4.47 $\pm$ 0.04 & -0.41 $\pm$ 0.02 & 0.45 & -4.96 & 5.0 / 5.2 \\
HD126793 & 0.92 $\pm$ 0.06 & 5910 $\pm$ 31 & 4.46 $\pm$ 0.03 & -0.71 $\pm$ 0.02 & 0.52 & -4.92 & 9.8 / 10.0 \\
HD175607 & 0.81 $\pm$ 0.05 & 5392 $\pm$ 17 & 4.51 $\pm$ 0.03 & -0.61 $\pm$ 0.01 & 0.70 & -4.92 & 28.9 / 29.7 \\

\end{tabular}
\tablefoot{
\tablefoottext{a}{The two values are estimated from the $\log R'_{HK}$ measurements  
                   using the calibrations from \citet{Noyes1984a} and \citet{Mamajek2008}, respectively.}
\tablefoottext{b}{After removing the observation at $BJD=2455304.48$ for which the $\log R'_{HK}$ value was not calculated.}
}
\end{table*}

For these 15 stars, we analyse a total of 1540 radial velocity measurements obtained using the HARPS spectrograph at the 3.6m ESO telescope (La Silla-Paranal Observatory).
All observations were reduced with the HARPS pipeline (version 3.5), where the radial velocities are obtained by performing a cross-correlation of the observed spectrum with a numerical mask \citep{Baranne1996,Pepe2002}.

For the first measurements (obtained before the present Large Program and still within the HARPS GTO program), the exposure times were not long enough to average out the noise coming from stellar oscillations and the radial velocities had a limiting precision of around $2\ms$. 
The observing strategy was since revised by setting exposure times to 900\,s, resulting in a precision of the order of $1\ms$. 
With the start of the Large Program, in October 2012, we aimed at obtaining more than one spectrum of the star in a given night, separated by several hours. 

We use nightly binned data in our analysis, thus focusing on signals with periods longer than one night. 
The objective of this strategy is to minimize the impact of granulation phenomena on the radial velocities and improve the planetary detection limits \citep[see][]{Dumusque2011b}. 
We note, however, that this observing strategy has some caveats since it has been designed for and tested in stars with solar metallicities and its optimality is not guaranteed for metal-poor stars. Also, considering the binned data means we cannot assess the presence of shorter period planets \citep[e.g.][]{Dawson2010,Hatzes2014,Grunblatt2015}.

\begin{table}
\caption{Number of individual night observations, 
         average RV errorbar, weighted standard deviation of the RV measurements 
         and timespan for each of the targets.}
\label{table:timeseries}
\centering          
\begin{tabular}{lcccc}
\hline
\hline

Star & N & $\bar{\sigma_i}$ & $s_{\text{RV}}$ & $\Delta t$\\
{\tiny } & {\tiny } & {\tiny [$\ms$]} & {\tiny [$\ms$]} & {\tiny [days]}\\
\hline
HD224817 & 100 & 1.15 & 1.98 & 4044 \\
HD21132 &  85 & 2.25 & 2.65 & 3935 \\
HD22879 &  80 & 0.81 & 1.60 & 4053 \\
HD31128 & 171 & 3.57 & 3.58 & 4081 \\
HD41248 & 160 & 1.26 & 3.41 & 3737 \\
HD56274 & 161 & 0.80 & 2.70 & 4046 \\
HD79601 &  79 & 1.00 & 1.77 & 3761 \\
HD87838 &  87 & 1.34 & 2.36 & 3983 \\
HD88725 &  90 & 0.79 & 2.35 & 3941 \\
HD111777 &  79 & 1.09 & 1.60 & 3756 \\
HD114076 &  77 & 1.41 & 1.87 & 3727 \\
HD119173 &  98 & 1.59 & 1.62 & 2959 \\
HD119949 &  81 & 1.50 & 2.23 & 3614 \\
HD126793 &  79 & 1.28 & 1.90 & 3595 \\
HD175607 & 113 & 1.08 & 2.62 & 3713 \\

\end{tabular}
\end{table}
\section{Methodology}\label{sec:method}

As a first step in the analysis of each star, and as is standard when searching for planets in RV data, we compute the generalised Lomb-Scargle (GLS) periodogram \citep{Zechmeister2009a} of the radial velocities and search for significant peaks. 
The significance of a peak is evaluated by the false alarm probability (FAP) estimated with a bootstrap permutation procedure, first devised by \citet{Murdoch1993}; see also \citet{Endl2001,Mortier2012}.
The RV measurements and errorbars are (together) randomly redistributed, with repetition, keeping the time-stamps of observations fixed. 
For each randomization, a new GLS periodogram is computed and its maximum power is compared with the original periodogram,
thus obtaining the FAP levels. 
After repeating this process 1000 times, we show the estimated 1\% and 0.1\% FAP levels.

If any periodogram peak is deemed significant, we proceed with the analysis of the RV data together with other activity proxies to try and determine the nature of the signal.
We will focus on the FWHM of the HARPS CCF and the \rhk proxy as the most reliable indicators of activity, even though we have also analysed other indicators.
It may also be that a signal is present in the data which was created by a planet with a moderately high eccentricity. 
The standard GLS periodogram (which searches for sinusoidal signals) is then not the best tool to recover it. 
Therefore, for the stars in which the periodogram does not show significant peaks, we still run a planet detection algorithm that compares the evidence for models with 0, 1 and 2 Keplerian signals (see below). 
For some stars which show evidence for a Keplerian signal, we further try to assert its nature.

As is now ubiquitous in the literature, our model for the radial-velocity observations $v_i$ is described by a Gaussian sampling distribution
\begin{equation}\label{eq:sampling-dist}
v_i \given M_n \sim \mathcal{N}(V_0 + V_n\,(t_i, \theta), \, \sigma_i^2+s^2)
\end{equation}
where $M_n$ represents a model with $n$ planets, $V_0$ is a constant offset, $V_n\,(t_i, \theta)$ is the radial-velocity shift caused by $n$ orbiting planets, which depends on time $t_i$ and the planets' orbital parameters $\theta$. The parameter $s$ is an extra ``jitter'' term, added in quadrature to the total measurement uncertainties $\sigma_i$, which allows for additional white noise to be taken into account. 

We assume each planet induces a Keplerian signal $k\,(t)$, such that $V_n\,(t) = \sum_n k\,(t)$. This signal depends on the orbital period $P$, semi-amplitude $K$, eccentricity $e$, argument of the periastron $\omega$ and time of periastron passage $T_0$ \citep[e.g.][chap. 2]{Perryman2014}. When $n=0$ we have $V_n=0$ by definition. 

To infer the number of Keplerians supported by a dataset $d=\{t_i, v_i, \sigma_i,\: i=1,\ldots,N\}$ we can calculate the marginal likelihood, or evidence $E_n$, for each of the models $M_n$:
\begin{equation}\label{eq:evidence}
E_n = \p[d \given M_n] = \int_\Theta \p[d \given \Theta, M_n] \p[\Theta \given M_n] d\Theta \,. 
\end{equation}
where $\p[d \given \Theta, M_n]$ is the likelihood function and $\p[\Theta \given M_n]$ the prior distribution for the parameters of the model. Note that $\Theta$ has different dimensions, depending on the value of $n$ and corresponds to the full set of parameters to be marginalised over, $\Theta=\{V_0, s, \theta\}$. The likelihood is given by the product of the terms in Eq. \eqref{eq:sampling-dist} for every data point,
\begin{align}
  \p[d \given \Theta, M_n] = \left(2\pi\right)^{-N/2} &
                             \left[ \prod_{i=1}^N \left(\sigma_i^2 + s^2\right)^{-1/2} \right] \times \nonumber \\ 
                           & \times \exp \left \{ - \sum_{i=1}^N \frac{\left[ v_i - V_0 - V_n\,(t_i, \theta) \right]^2}{2\left(\sigma_i^2 + s^2\right)} \right\}
\end{align}

The comparison between any pair of models is made by evaluating the associated odds ratio \citep[e.g.][]{Gregory2010}
\begin{equation}\label{eq:odds}
	\mathcal{O}_{ij} = \frac{\p[M_i \given d]}{p\,(M_j \given d)} = \frac{\p[M_i]}{p\,(M_j)} \cdot \frac{\p[d \given M_i]}{p\,(d \given M_j)}
\end{equation}
which simplifies to the ratio of the evidences when there is no prior preference for any model, that is, when $\tfrac{p\,(M_i)}{p\,(M_j)}=1$. If there is a preference for a given model, it can be included directly in Eq. \ref{eq:odds}. In this work we chose to consider all models equally probable \emph{a priori}. \citet{Kass1995} provide a qualitative scale for the interpretation of evidence ratio values.

To evaluate the difficult (multidimensional) integral in Eq. \eqref{eq:evidence}, we use the nested sampling algorithm \citep{Skilling2004} implemented in {\sc MultiNest} \citep{Feroz2009,Feroz2013}. 
The key idea in nested sampling is to update a set of points, originally sampled from the prior, with new samples that are subject to a hard likelihood constraint. 
At each iteration, the algorithm progressively moves towards regions of higher likelihood. Each time a new sample is obtained, it is assigned a value $X \in [0, 1]$ representing the amount of prior mass estimated to lie at a higher likelihood than that of the discarded point\footnote{The prior mass element is $dX=\p[\Theta \given M_n] d\Theta$ such that $E=\int L\,dX$, with $L$ the likelihood function.}.
Assigning these X-value maps from the parameter space to $[0, 1]$ or, in other words, divides the prior volume into a large number of points with equal prior mass. In the new space, the prior becomes a uniform distribution and the likelihood is a decreasing function of $X$. Then, the evidence can be computed by simple quadrature. For further details, the reader is referred to \citet{Skilling2004,Mukherjee2006,Sivia2006,Feroz2009}.

Nested sampling also provides posterior samples for all parameters as a by-product. The necessary descriptive statistics can then be calculated from these samples. Estimating all the posterior distributions for a particular model is what we mean by \emph{fitting} a Keplerian in the remainder of the paper. This algorithm has been tested and validated in the analysis of radial-velocity data by \citet{Feroz2011a,Feroz2011} and \citet{Feroz2013a}.

Another way to approach the problem is to treat $n$ (the number of Keplerians) itself as a free parameter of the model (so that it is part of $\Theta$) and sample from its own posterior distribution, $\p[n \given d]$, to infer the number of Keplerians present in the data. We use the recent algorithm proposed by \citet{Brewer2015} to sample from this distribution. The method uses trans-dimensional birth-death moves, within the context of diffusive nested sampling \citep{Brewer2011}, and was found to give results comparable to {\sc MultiNest}, often with a smaller number of likelihood evaluations and less computation time. The same likelihood function and priors were used for both methods. 

We further need to specify the priors for each of the parameters. 
It is through the priors that the Bayesian analysis can take into account the principle of parsimony and Occam's razor. What is important is not (only) the \emph{number} of parameters in a model but instead the amount of support in their priors. This is the reason to choose uninformative and physically motivated priors. 

The form and limits of the priors we use are shown in Table \ref{tab:priors}. A few of our choices merit further discussion. 
For the orbital periods, we use a Jeffrey's prior between 1d and the timespan of each dataset. Apart from HD56274 and HD175607, which are analysed separately, none of the stars show clear long-term trends that could suggest the presence of long period planets. We thus limit our search for periods shorter than the timespans, which are in fact very long already. The lower limit is due to our use of nightly binned observations. 
For the semi-amplitudes, a modified Jeffrey's prior is used, spanning the range $0-10\ms$ and with a ``knee'' at the mean error bar of each dataset. The choice of upper limit comes from the cuts used when defining the complete metal-poor sample \citep[see][]{Santos2014}, which was stripped of stars showing large RV dispersions\footnote{In our full sample, the highest radial-velocity rms is $4.1\ms$.}. 
The same arguments apply to the jitter prior.
The prior for the orbital eccentricities was suggested by \citet{Kipping2013} after an analysis of the population of known exoplanets.
Priors for the other parameters are uniform between what we consider to be sensible limits.

Model selection criteria other than the evidence can be more easily calculated using just the maximum-likelihood value $\hat{\rule{0ex}{1.6ex} L}$ and the total number of parameters in the model, $N_\text{par}$. 
When presenting our results, in Table \ref{table:model-selection}, we include the values of the sample-corrected Akaike Information Criteria (AIC$_C$) and the Bayesian Information Criteria (BIC), calculated as described in \citet{Burnham2010}:
\begin{equation*}
\begin{split}
  AIC_C & = -2 \ln \hat{\rule{0ex}{1.6ex} L} + 2\,N_\text{par} + \frac{2K(N_\text{par}+1)}{n-N_\text{par}-1} \\
  BIC & = -2 \ln \hat{\rule{0ex}{1.6ex} L} + N_\text{par} \ln N
\end{split}
\end{equation*}
The preferred model, when using these criteria, is the one which gives the lowest AIC$_C$ and BIC values. 
For the choice of one model to be significant one can require differences $\Delta$AIC$_C$ and $\Delta$BIC around 10 \citep{Burnham2010}.

\begin{table}
\caption{Priors for the model parameters.}
\label{tab:priors}
\centering    
\begin{tabular}{lc}
\hline\hline
$P$ [days]        & $\mathcal{J}\,(1 \,, \text{timespan})$          \\
$K$ [$\ms$]       & $\mathcal{MJ}\,(\bar{\sigma_i}, 10)$            \\
$e$               & $\mathcal{B}\,(0.867, 3.03)$                    \\
$\omega$ [rad]  & $\mathcal{U}\,(0, 2\pi)$                          \\
$T_0$ [days]      & $\mathcal{U}\,(t_0 \,, t_0 + \text{timespan})$  \\[1ex]
\hline          
$V_0$ [$\ms$]      & $\mathcal{U}\,(\min v_i\,, \max v_i)$ \\ 
$s$ [$\ms$]      & $\mathcal{MJ}\,(\bar{\sigma_i}, 10)$ \\
\end{tabular}
\tablefoot{The average uncertainty $\bar{\sigma_i}$ is used in the priors for the semi-amplitude and jitter.
           The symbols have the following meaning: 
           $\mathcal{J}(\cdot, \cdot)$ -- Jeffreys prior with lower and upper limits; 
           $\mathcal{MJ}(\cdot, \cdot)$ -- Modified Jeffreys prior with knee and upper limit;
           $\mathcal{B}(\alpha, \beta)$ -- Beta prior with shape parameters $\alpha$ and $\beta$;
           $\mathcal{U}(\cdot, \cdot)$ -- Uniform prior with lower and upper limits;
           See, e.g., \citet{Gregory2007,Kipping2013} for the mathematical forms of each distribution.
           }
\end{table}

\section{Case-by-case results}\label{sec:results}

In this section we present the individual analysis for each star. 
Plots with the radial-velocity data and GLS periodograms are shown in Figs. \ref{fig:HD87838}-\ref{fig:HD126793} and a summary of the model selection results, for the stars without significant periodogram peaks, is shown in Table \ref{table:model-selection}.


\subsection{Candidate planetary signals}\label{sec:unconfirmed-signals}

\paragraph{HD87838} 
  This star was observed 104 times on 87 different nights. 
  The radial velocities show a dispersion of $2.4\ms$ and a mean uncertainty of $1.3\ms$. 
  The GLS periodogram shows no significant peaks (Fig. \ref{fig:HD87838}) but the Bayesian analysis finds slight evidence for the presence of one Keplerian signal (Table \ref{table:model-selection}).

\begin{figure}
  \centering
  \includegraphics[width=\hsize]{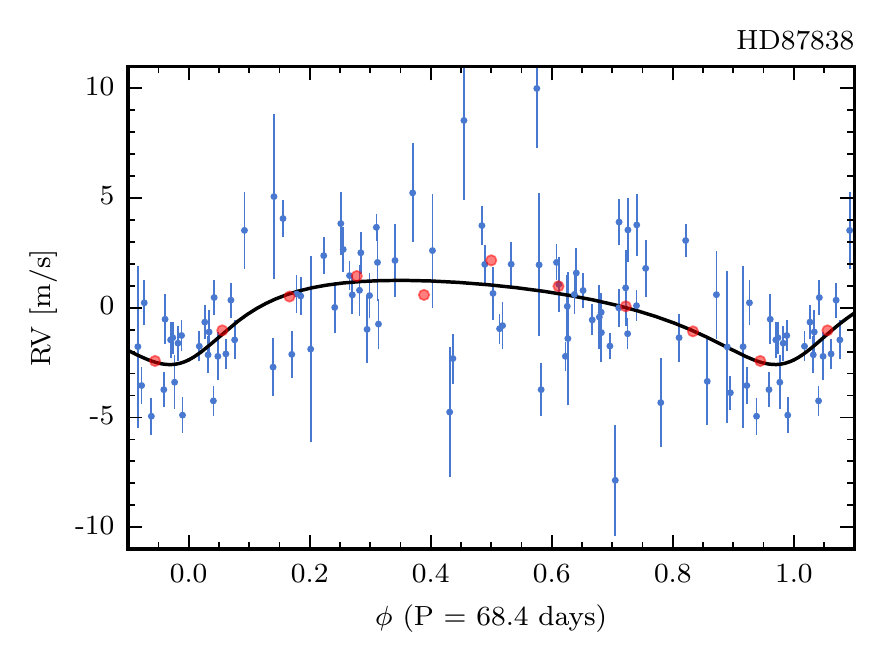}
    \caption{Phase-folded RV curve for HD87838, with a period of 68.42\,d. The blue points are the measured radial-velocities, red circles are binned in phase, with a bin size of 0.1.}
  \label{fig:HD87838_fit_phase}   
  \end{figure}

  The strongest peaks in the periodogram are at 988\,d, 68\,d and 334\,d. 
  The most convincing (highest likelihood) Keplerian fit is a signal with 
  $P=68.34 \pm 0.46$\,d, 
  a semi-amplitude $K=1.9 \pm 0.38 \ms$ 
  and an eccentricity of $0.46 \pm 0.16$. 
  If caused by a planet, this signal would correspond to a planet with 
  a minimum mass of $11.14\pm2.47 M_\oplus$ 
  at an orbital distance of $0.322 \pm 0.001$ AU. 
  The phase-folded radial-velocities for this solution are shown in Fig. \ref{fig:HD87838_fit_phase}.

\begin{figure}
  \centering
  \includegraphics[width=\hsize]{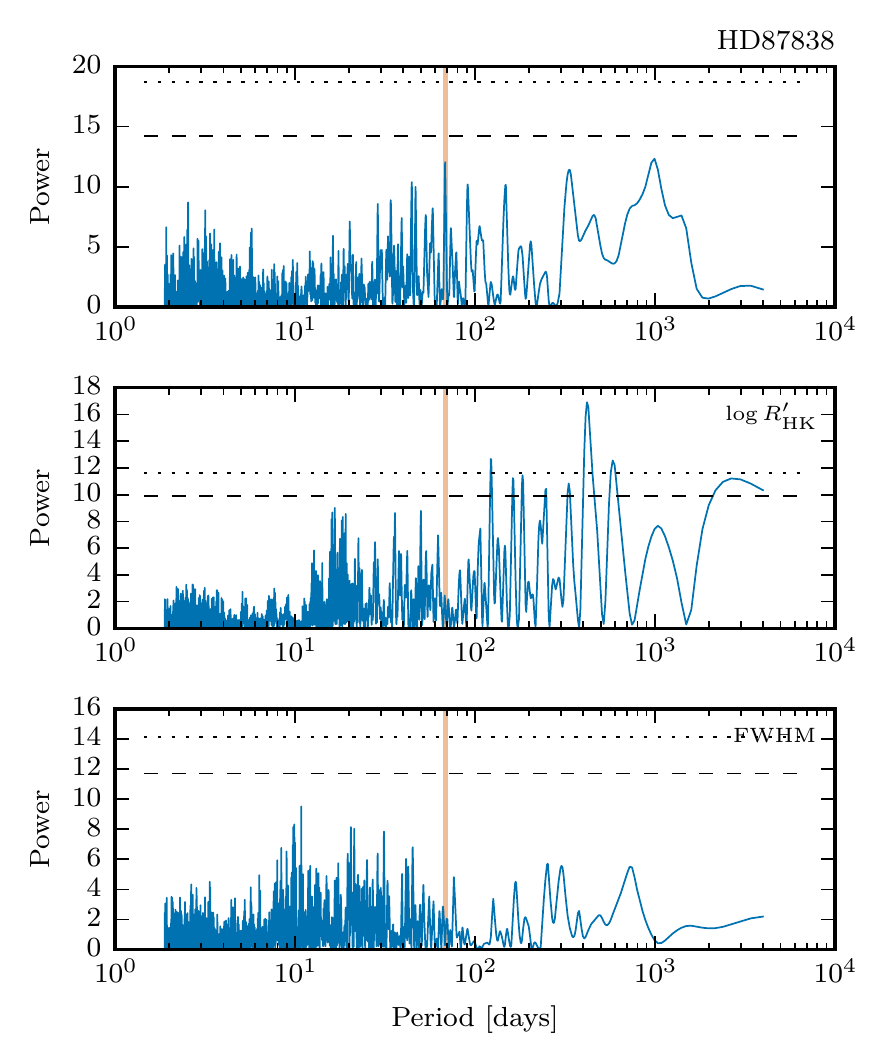}
    \caption{Periodograms of the radial velocities (top) the \rhk (middle) and the FWHM of the CCF for HD87838. The vertical line shows the orbital period of the candidate planet, at 68.3\,d.}
  \label{fig:HD87838_indicators}   
  \end{figure}

  Analysing the \rhk and FWHM activity indicators, we do not find strong periodicities close to 68\,d (Fig. \ref{fig:HD87838_indicators}), and therefore it is unlikely that this signal is caused by stellar activity (indeed, the periodogram of the FWHM shows a non-significant peak close to the estimated rotation period). Since the Bayesian analysis only gives marginal evidence for a Keplerian signal (a Bayes factor of 1.75 is ``not worth more than a bare mention'' in the scale of \citealt{Kass1995}), we consider that this detection is marginal and needs further confirmation. 
  More data would certainly shed light on its true origin, and we will continue to observe this star in the future.

\paragraph{HD175607} 
  This star's periodogram (Fig. \ref{fig:HD175607}) shows a very significant peak at 29\,d, coinciding almost exactly with the estimated rotation period (see Table \ref{table:parameters}). Some of the longer periods found in the RVs are also present in the $\log R'_{HK}$ indicator, but no indication of the 29\,d period is found in any of the activity proxies. A full analysis of these data is presented in \cite{Mortier2016} where the authors find evidence for (at least) one Neptune-mass planetary companion. In Sect. \ref{sec:detection-limits}, we consider the detection limits after having removed this orbital solution.


\subsection{Signals caused by stellar activity}\label{sec:signals-activity}

The effects of stellar activity on radial-velocity observations can be broadly divided in short-term effects, modulated by the stellar rotation, and long-term effects caused by global activity cycles \citep{Baliunas1995}. 
The activity-induced signals can be diagnosed using activity proxies like line-profile indicators \citep{Queloz2001,Dumusque2011,Boisse2011} or the \rhk values \citep[e.g.][]{Bonfils2007}.
One way to correct the radial velocities for the effects of stellar activity is to assume a linear correlation between, e.g., the \rhk activity index and the activity-related radial-velocity variations and remove this correlation from the RVs \citep[e.g.][]{Melo2007}. Assuming an estimate of the rotation period of the star, one can also subtract sinusoidal functions from the RVs, at the rotation period and its harmonics \citep[e.g.][]{Boisse2011}.
The following four stars show evidence for these activity-induced signals.

\paragraph{HD21132} 
  A total of 85 (107 before binning) radial-velocity measurements were obtained for this star, spanning a baseline of over 10 years. 
  The average error bar and dispersion are relatively high, probably due to the high effective temperature (implying a higher oscillation and granulation noise, \citealt{Dumusque2011a}). 
  All the peaks in the periodogram are below the 1\% FAP (Fig. \ref{fig:HD21132}). 
  The model selection analysis nevertheless finds positive evidence for one Keplerian signal, with the best solution having a period of 3712\,d and a 4.9$\ms$ amplitude.

\begin{figure}
  \centering
  \includegraphics[width=\hsize]{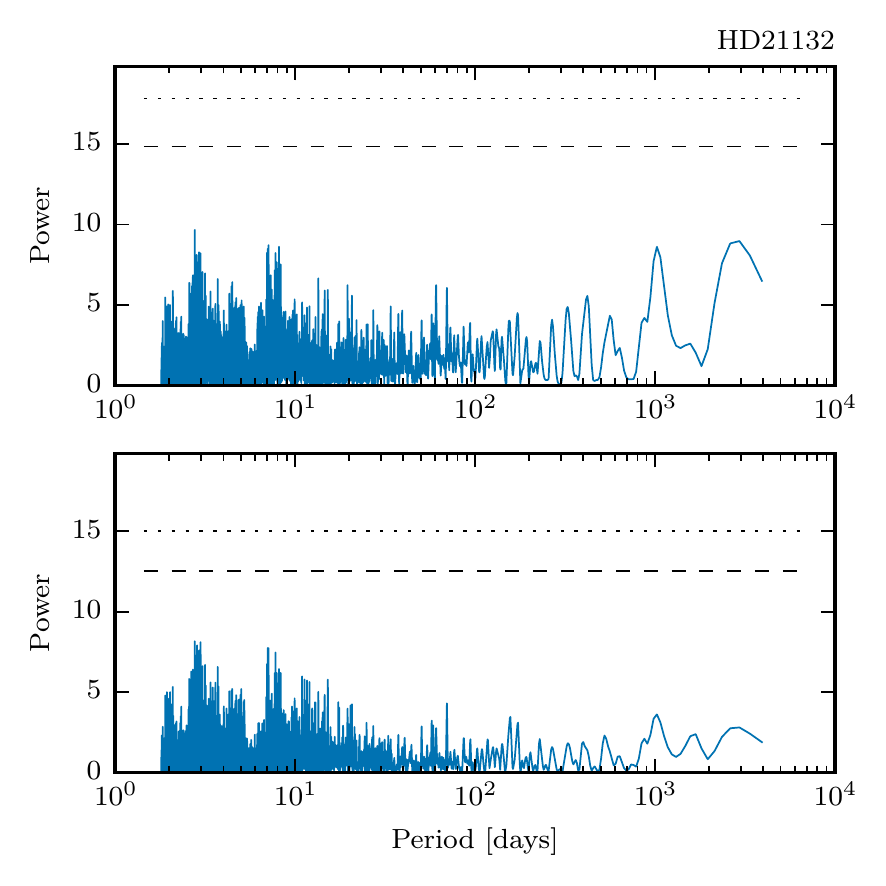}
    \caption{Periodograms of the radial-velocities of HD21132, before (top) and after (bottom) correcting for a linear variation with \rhk. }
       \label{fig:HD21132-per-after-rhk-linear}
  \end{figure}

  The very long period of this solution (the timespan of observations is 3935\,d) leads us to believe that it is not of planetary origin.
  In fact, some of peaks at long periods, seen in the periodogram of the RVs, are also found in the periodogram of the \rhk. 
  When the RVs are corrected with a linear variation on this indicator, these peaks show a clear decrease in power (Fig. \ref{fig:HD21132-per-after-rhk-linear}). 
  No additional signals were found after applying this correction. We thus conclude that the marginally significant signal found in the data of HD21132 is probably best explained by a long-term magnetic cycle with a period around 10 years.

\paragraph{HD41248} 
  This star was the subject of a recent debate regarding the presence of a planetary system composed of two super-Earths. 
  From an analysis of the first $\sim\!60$ measurements of this dataset, \citet{Jenkins2013} found evidence for the presence of two planets, at orbital periods of 18.357\,d and 25.648\,d. 
  Adding about 160 new radial-velocities, \citet{Santos2014} attributed one of the signals (25\,d) to activity and could not recover convincing evidence for the second planet. 
  More recently, \citet{Jenkins2014} re-analysed the new data and again found the two planet model to be the most probable.
  No new measurements were made for this star following the latter two works, but we include it here for completeness. 

\begin{figure}
  \centering
  \includegraphics[width=\hsize]{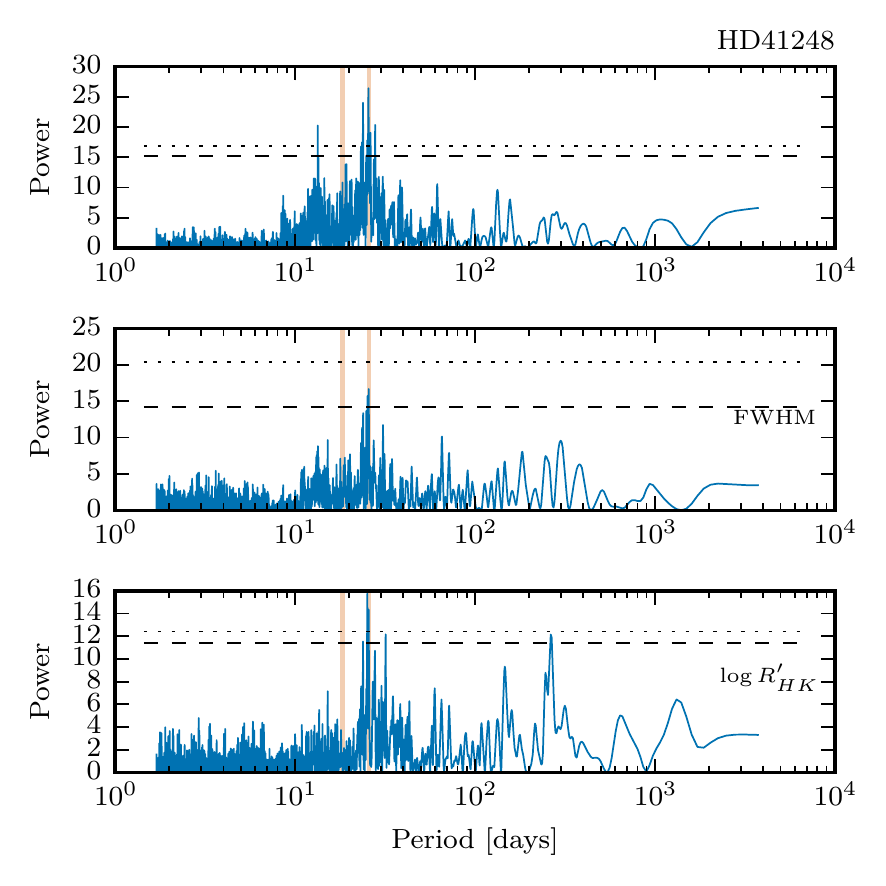}
    \caption{Periodograms of the radial-velocities, FWHM and \rhk for HD41248. The vertical lines show the periods of the planets announced by \citet{Jenkins2013}.}
       \label{fig:HD41248-per-activity}
  \end{figure}

  The period at 25\,d is found to be significant, both in the periodogram and from the model selection analysis (with an odds ratio\footnote{The odds ratio cannot be compared to the one found by \citet{Jenkins2014} because we did not use the same model for the RVs.} of $\sim\!6\times10^4$). 
  This would be very strong evidence in favour of (at least one) planetary companion. 
  However, both the FWHM of the HARPS CCF and the \rhk show significant peaks at the same period (Fig. \ref{fig:HD41248-per-activity}). 
  In light of this evidence we must consider this signal to be caused by stellar activity \citep[see also][]{SuarezMascareno2015}.

  The 18\,d peak is not recovered clearly in the full dataset. 
  In addition, a convincing detection of this planet presupposes a correction for the activity-induced signal.
  When attempting to perform this correction, by removing sinusoids at 25\,d and its harmonics, we do not find convincing evidence for the 1-planet model.
  The orbital period of the reported planet, as detected by \citet{Jenkins2014}, is also very close to our estimate of the rotation period. 
  It is clear that the currently available data do not allow a firm conclusion on the presence of a planetary system around HD41248.

\begin{figure}
  \centering
  \includegraphics[width=\hsize]{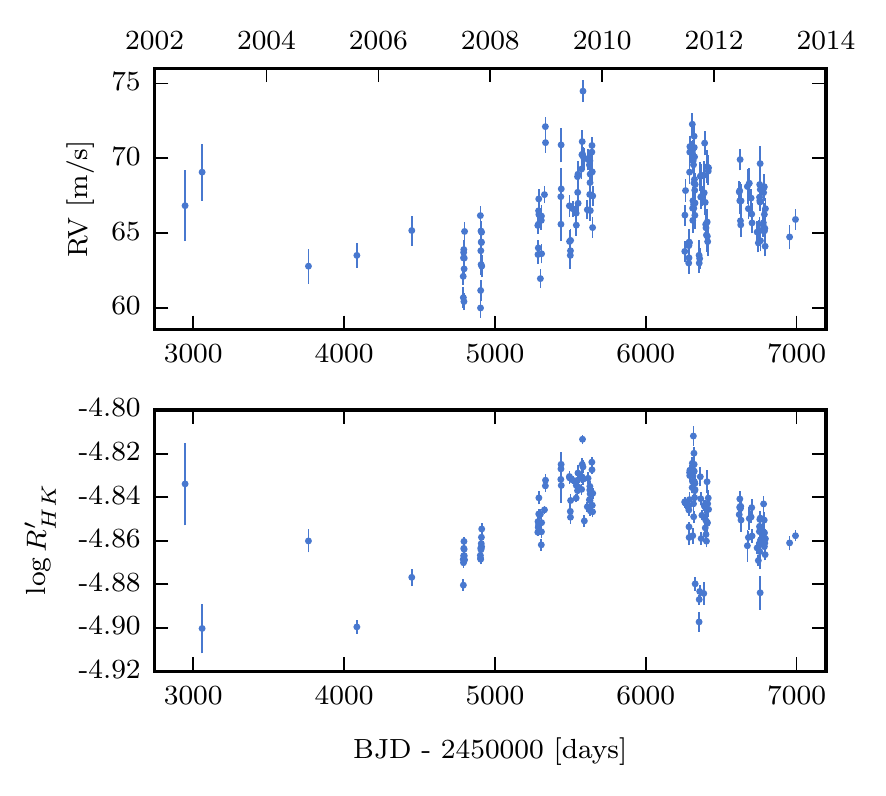}
    \caption{Time series of the radial velocities (top) and \rhk activity index (bottom) for HD56274.}
       \label{fig:HD56274-rv-and-rhk}
  \end{figure}

\begin{figure}
  \centering
  \includegraphics[width=\hsize]{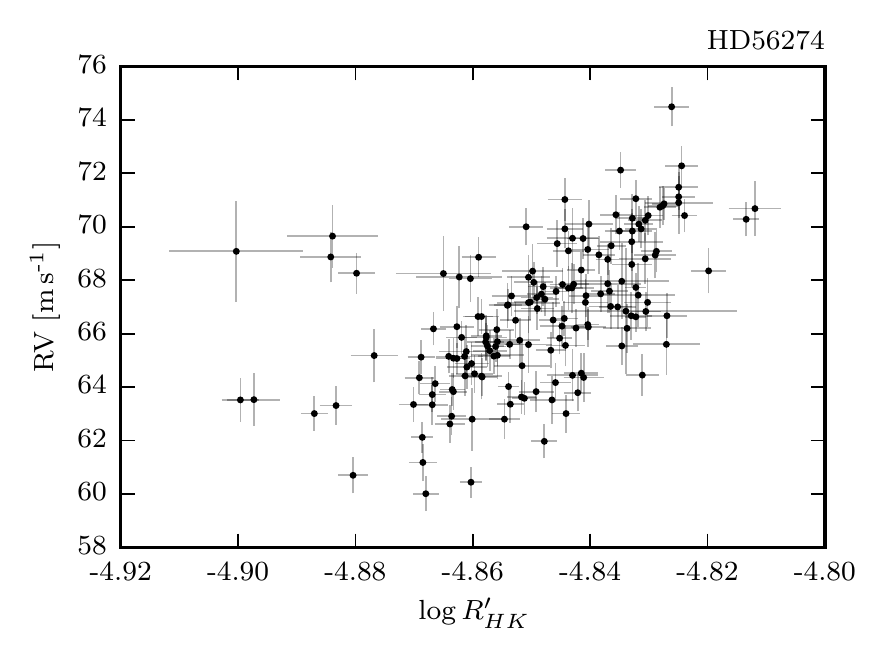}
    \caption{Radial-velocity versus the \rhk activity indicator for HD56274. We estimate the posterior distribution for Spearman's rank correlation coefficient using an MCMC. The posterior has a mean value of 0.62 and a 95\% credible interval $[0.52, 0.71]$ for these data, suggesting a tight correlation, likely caused by a long-term activity cycle.}
       \label{fig:HD56274-rv-vs-rhk}
  \end{figure}

\paragraph{HD56274}
  This star was followed closely for over 11 years, with 228 observations in 161 nights. 
  The RVs show a clear long term modulation (Fig. \ref{fig:HD56274-rv-and-rhk}), which is also evident from the periodogram (Fig. \ref{fig:HD56274}). 
  This modulation is correlated with the \rhk indicator (Fig. \ref{fig:HD56274-rv-vs-rhk}) which shows a clear long-term magnetic cycle. 
  The periodogram of the \rhk values suggests a period of over 10 years for this activity cycle. 

  From the relations of \citet{Noyes1984} and \citet{Mamajek2008}, we estimate the rotation period of HD56274 to be around 13\,d (see Table \ref{table:parameters}). 
  A peak at this period appears in the periodogram of the RV data, slightly above the 1\% FAP line, but not in the observed values of the activity indicators. 
  Interestingly, when removing a cubic polynomial from the \rhk values (as a correction for the magnetic cycle) a peak at 14.5\,d appears in the periodogram of the residuals, although not at a significant level. 
  This suggests that the rotation period of the star is indeed around 14\,d such that this signal is present in both RVs and the \rhk.

\begin{figure}
  \centering
  \includegraphics[width=\hsize]{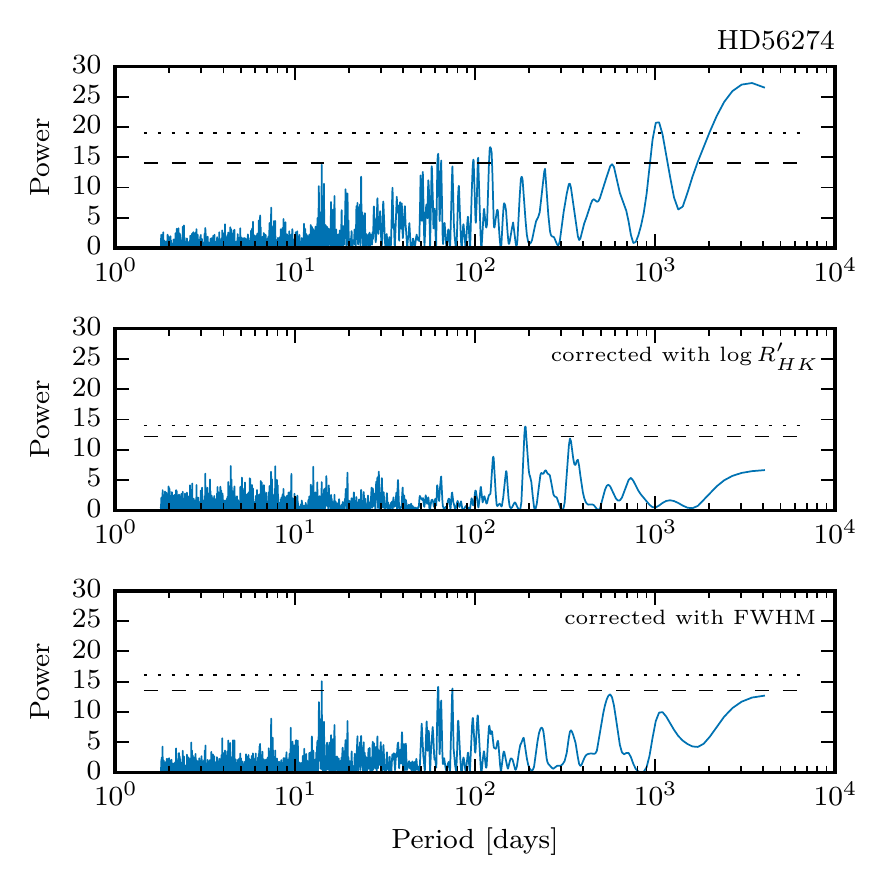}
    \caption{The top panel shows the GLS periodogram of the observed radial-velocities of HD56274. The middle and bottom panels show the periodograms after correcting the RVs from linear correlations with the \rhk activity indicator and the FWHM, respectively.}
       \label{fig:HD56274-multiple-pers}
  \end{figure}

  To try to correct the RVs for the long-term activity-induced signal, we removed a linear correlation both with \rhk and with FWHM (which also correlates with the RVs). 
  The periodograms of the corrected RVs are shown in Fig. \ref{fig:HD56274-multiple-pers}.
  Correcting with \rhk removes most of the long period peaks but leaves in the periodogram the peaks associated with one year aliases.
  The cause for these peaks is probably the same instrumental effect identified in the case of HD88725 (see Sect. \ref{sec:other-cases}).
  No other significant signals can be found in the residual data.
  The correction with the FWHM reveals a significant peak at 14\,d (close to the estimated rotation period) and two peaks at 62\,d and 74\,d. 
  With this second correction, however, the long-term periodicities are not completely removed.
  We conclude that with the available data, and after correcting for the long-term magnetic cycle, it is not possible to assert the presence of planets orbiting this star.

\begin{figure}
\centering
\includegraphics[width=\hsize]{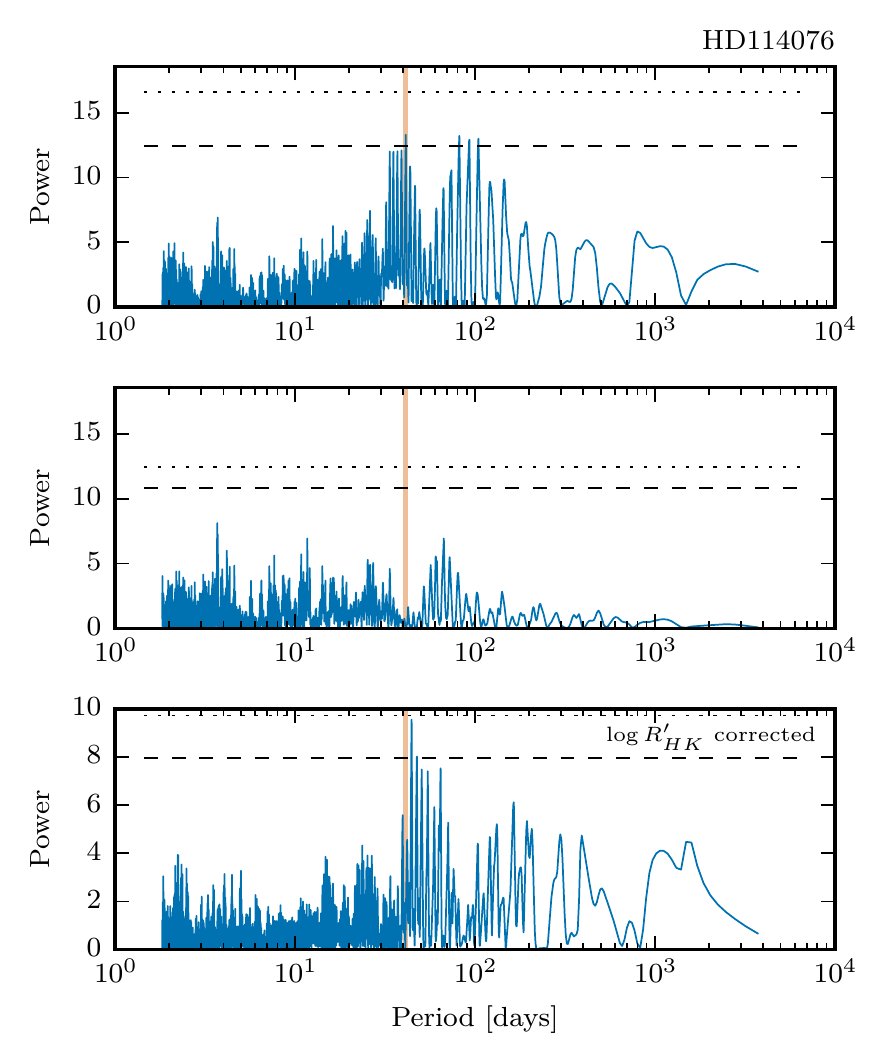}
  \caption{Periodogram for HD114076 before (top) and after (middle) fitting and removing a sine function to correct for activity-induced variations. The vertical dashed line marks the best fit period of 41.27\,d. The bottom plot shows the periodogram of the \rhk index after removing a quadratic polynomial from the observed values.}
     \label{fig:HD114076-per}
\end{figure}

\paragraph{HD114076} 
  The rotation period for this star is estimated to be between 41 and 43\,d, depending on the calibration (Table \ref{table:parameters}). 
  The periodogram (Fig. \ref{fig:HD114076-per} top panel, and Fig. \ref{fig:HD114076}) shows a cluster of peaks around 40\,d and also around 80\,d. 
  The strongest peak is at 41.3\,d and very close to the 1\% FAP.

  To correct the RVs from what we believe is an activity-induced signal, a sine function was fitted and subtracted from the data.
  The best fit period was 41.27\,d, and the periodogram of the residuals, shown in Fig. \ref{fig:HD114076-per} (bottom panel), does not present additional significant signals. 

  To confirm that this signal is caused by the rotation of the star, we analysed the activity indicators.
  In particular, the \rhk shows a clear long-term trend, which might be caused by a magnetic cycle.
  We tried to remove the long-term variations by fitting both quadratic and cubic polynomials to the \rhk observations.
  The residuals of both fits show a clear periodic signal at 44.4\,d which helps to corroborate the activity-induced nature of the 41\,d signal in the RVs.

\subsection{Sampling and instrumental effects}\label{sec:other-cases}

One additional obstacle in the detection of periodic RV variations due to planets is the presence of spurious signals and periodicities caused by the discrete time sampling of the observations or by instrumental effects. These can originate, amongst other effects, from Earth's rotation and orbital motion. The following stars show evidence of contaminations of this kind.

\paragraph{HD22879} 
  Although the periodogram of this star (Fig. \ref{fig:HD22879}) does not show significant peaks, the Bayesian analysis finds evidence for the presence of one Keplerian signal with a period of $\sim$770\,d and an eccentricity of 0.7. 
  This corresponds to the second highest peak in the periodogram.

\begin{figure}
  \centering
  \includegraphics[width=\hsize]{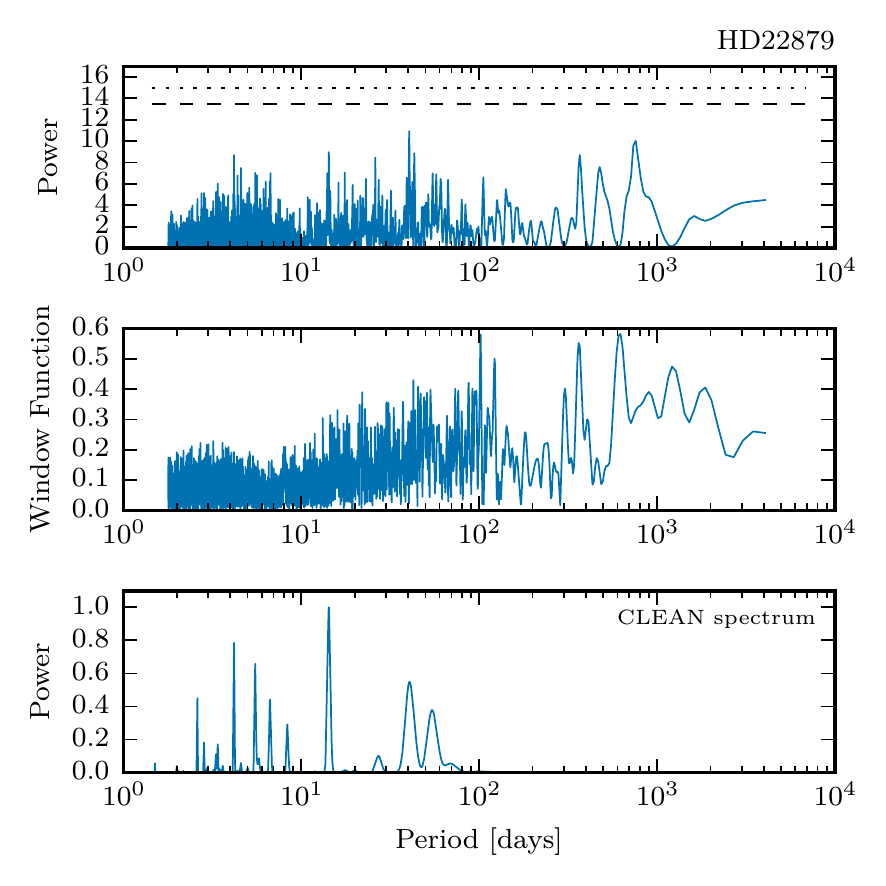}
    \caption{Periodogram, window function and CLEAN spectrum for the radial-velocities of HD22879.}
       \label{fig:HD22879-per-clean}
  \end{figure}

  We find that this signal can, however, be caused by the time sampling of the observations. In the top and middle panels of Fig. \ref{fig:HD22879-per-clean}, we show the periodogram together with the window function of the observations. 
  The window function is calculated as the Fourier transform of the observation times \citep{Roberts1987,Dawson2010} and provides information on the power that is introduced in the periodogram due to the sampling. 

  The {\sc CLEAN} algorithm, introduced for frequency analysis by \citet{Roberts1987}, is a well-known method to deconvolve the observed spectrum and the window function, thereby reducing the artifacts introduced by the sampling. Applying the {\sc CLEAN} algorithm to the RV measurements of HD22879, results in the spectrum shown in the bottom panel of Fig. \ref{fig:HD22879-per-clean}. The highest peak is at 14\,d, which is close to the estimated stellar rotation period. It is not clear if this power is activity-induced, since the activity indicators do not show any clear signal at this period (nor any long-term drifts).

  We conclude that the Keplerian signal found at 770\,d is best explained as originating from the time sampling. It is worth mentioning here that the Bayesian analysis is vulnerable to this kind of signals because it does not take into consideration, at any stage, the information present in the window function.

\begin{figure}
  \centering
  \includegraphics[width=\hsize]{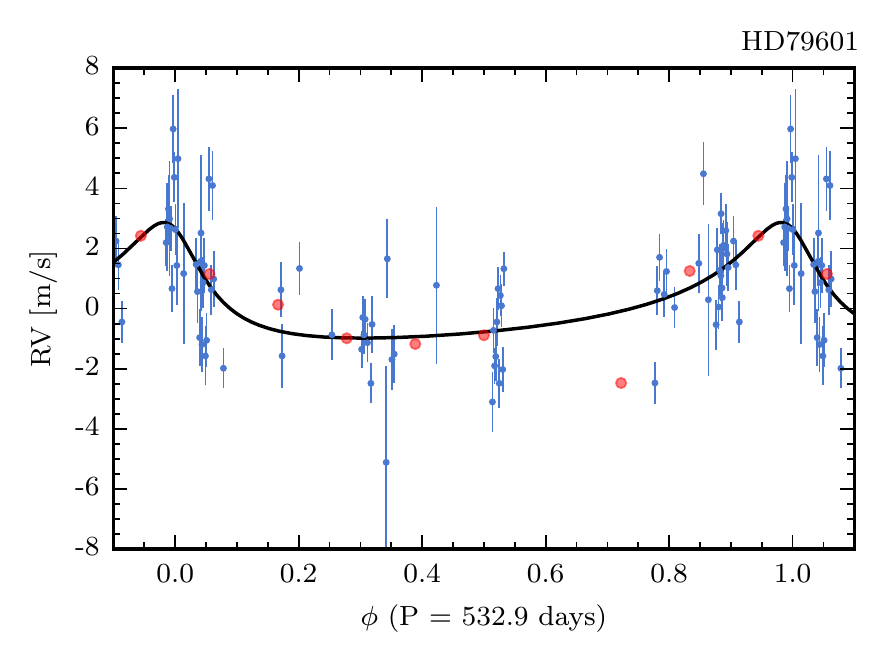}
    \caption{Phase folded radial-velocity measurements (blue) of HD79601 with a period of 532.9 days. The red circles are binned in phase, with a bin size of 0.1, and the black line shows the best fit Keplerian function.}
       \label{fig:HD79601-fit}
  \end{figure}

\paragraph{HD79601} 
  The periodogram of the radial-velocities of this star shows one significant peak at $\sim\!\!550$\,d (Fig. \ref{fig:HD79601}).
  A Keplerian fit at this period results in a solution with
  $P=532.85 \pm 9.63$\,d, 
  a semi-amplitude of $1.92 \pm 0.28 \ms$ 
  and an eccentricity of $0.55 \pm 0.18$. 
  A planet with these parameters would have a mass of $20.4 \pm 4.16 M_\oplus$.
  The phase-folded radial-velocities for this solution (Fig. \ref{fig:HD79601-fit}) show that the phase coverage is not ideal. The residuals from this fit do not show any significant periodogram peaks or evidence for additional signals.

  The lack of a good phase coverage of the orbital solution, puts the planetary hypothesis in question. From Fig. \ref{fig:HD79601} it is clear that the measurements were obtained in several series, separated by large time gaps. Looking only at the observations after BJD$=2454500$, the data show hints for a linear drift. When a linear drift is removed from this subset of the data (and also when a quadratic drift is removed from the full dataset), the peak at 550\,d vanishes and no other significant peaks remain in the periodogram of the residuals. At this point, our data do not allow us to reach any firm conclusions about the nature of the signal observed for HD79601.

\paragraph{HD88725} 
  The radial velocities of this star show a significant periodicity at 368.3\,d (Fig. \ref{fig:HD88725}).
  Because this period is close to one year, we measured the correlation between the radial-velocity and the barycentric Earth radial velocity (BERV). 
  The posterior distribution for Spearman's rank correlation coefficient (which was sampled using MCMC; see Figueira et al., submitted) has a mean of -0.29 and a 95\% credible interval (the highest posterior density interval) is $[-0.46, -0.13]$.
  This anti-correlation suggests that the signal detected in the RVs might be induced by the orbit of the Earth around the Sun. 

  \citet{Dumusque2015} recently described the cause for signals of this type to show up in HARPS radial-velocities, and suggested a method to correct for this effect: the spectral lines that cross the so-called block stitchings in the HARPS CCD can be identified in the correlation mask used to derive the RVs. 
  We created two different correlation masks, one with only the lines falling close to the block stitchings, and one without those lines. 

\begin{figure}
  \centering
  \includegraphics[width=\hsize]{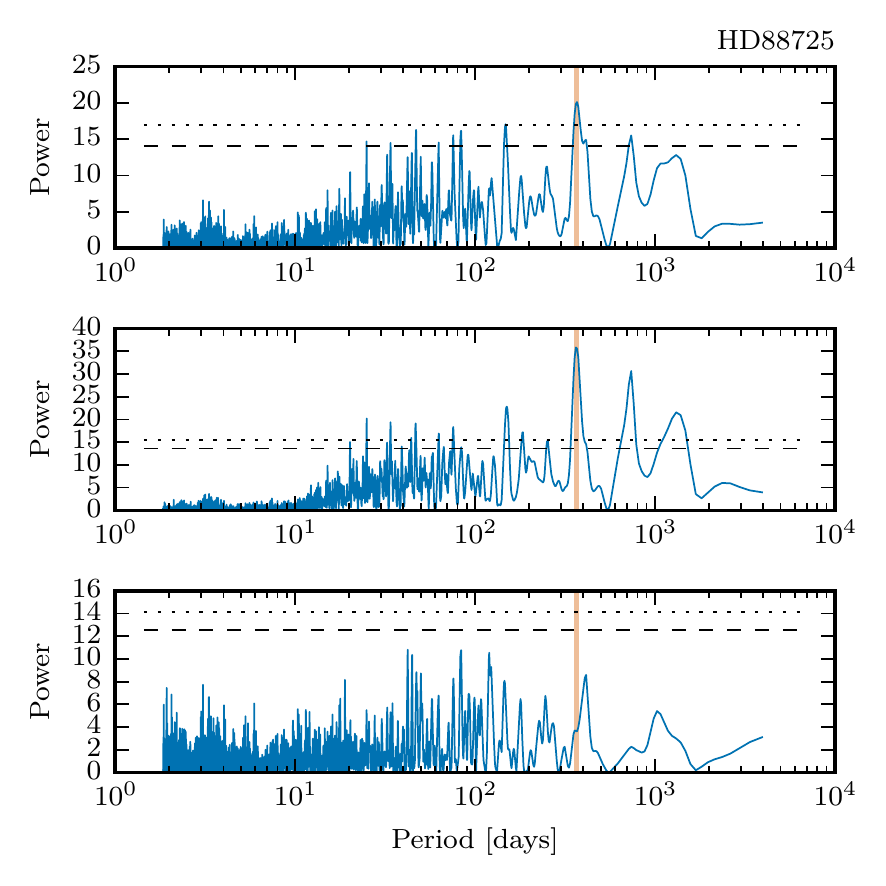}
    \caption{Periodograms of the original radial-velocities of HD88725 (top panel), the radial-velocities derived using a correlation mask that contains only the spectral lines crossing the CCD block stitchings (middle panel), and the radial-velocities derived using a correlation mask without those lines (bottom panel). The vertical line shows the orbital period of the Earth.}
       \label{fig:HD88725-gaps}
  \end{figure}

  For the case of HD88725, the periodograms of the RVs derived with the original mask and both custom masks are shown in Fig. \ref{fig:HD88725-gaps}. 
  The spectral lines that cross the block stitching are the origin of the peak at one year (middle panel) and when those lines are removed from the correlation mask, the peak vanishes from the periodogram (bottom panel). 
  After the RVs are corrected, the periodogram does not show significant peaks and we do not find evidence for the presence of Keplerian signals.

\begin{figure}
  \centering
  \includegraphics[width=\hsize]{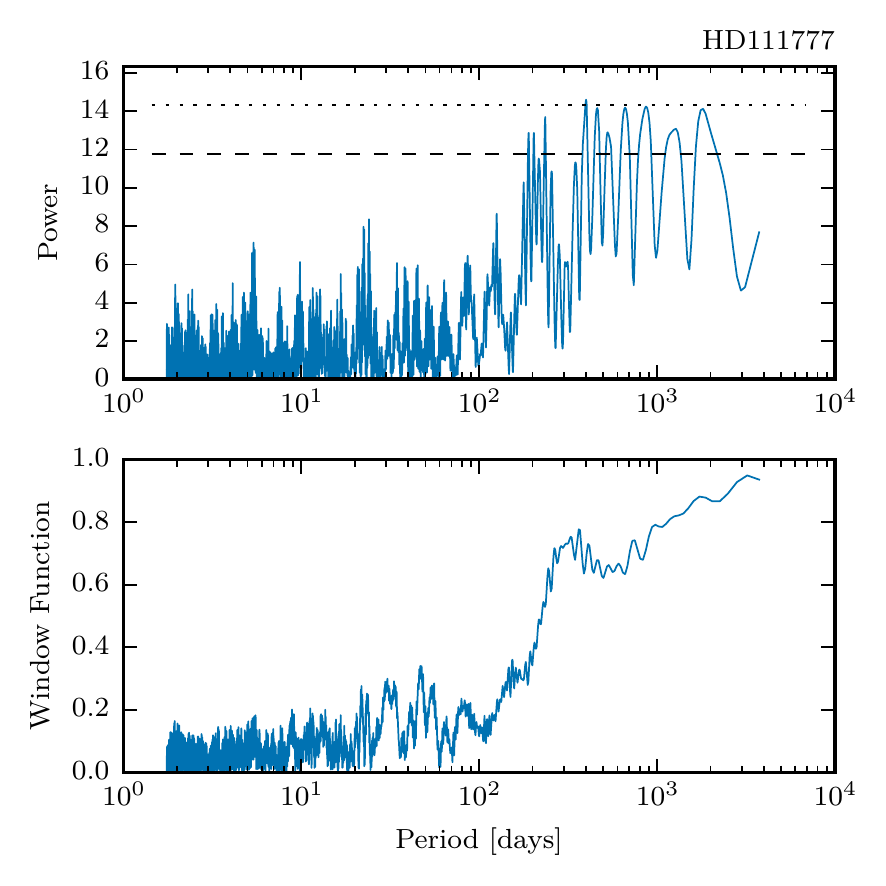}
    \caption{Periodogram and window function for HD111777.}
       \label{fig:HD111777-per}
  \end{figure}

\paragraph{HD111777} 
  In Fig. \ref{fig:HD111777-per}, the periodogram of the 79 measurements of this star is plotted together with the spectral window function (see also Fig. \ref{fig:HD111777}).
  Although there are significant peaks in the periodogram, it becomes impossible to identify clear peaks produced by physical signals, because of the confusion introduced by the sampling. 
  This does not mean that a long period signal is not present, only that the highest peak in the periodogram can show up in a different location due to the sampling. 

  Since the window function shows only excess power for long periods, and not distinct peaks, the CLEAN algorithm does not provide a good correction in this case. The large gap without observations is what causes these issues. After considering only the better-sampled subset of the data (after BJD=$2456000$), we do not find any significant signals. Also, after removing a linear trend\footnote{For consistency, we checked that the evidence of the constant model is higher than that of a model with an added linear drift parameter.} from the full dataset, the periodogram of the residuals does not show any significant peaks.

  The periodograms of the activity indicators do not help in corroborating the presence of a planetary signal: the FWHM shows a non-significant periodicity close to 730 days but the highest peak in the periodogram of the \rhk is at 16 days.
  We conclude, for the case of HD111777, that we are currently not able to assess the presence of a signal that might originate from a long-period planetary companion.

\paragraph{HD224817}
  The periodogram of the RVs of this star (Fig. \ref{fig:HD224817}) does not show peaks above the 1\% FAP line, but the model selection analysis finds the model with one Keplerian to be the most probable. In the scale of \citet{Kass1995}, an odds ratio of 1.25 (Table \ref{table:model-selection}) is ``not worth more than a bare mention''.

  The highest peak in the periodogram is at $\sim$245\,d -- the best fit Keplerian also converges to this solution -- but this period is probably caused by the deformation of the spectral lines that cross the block stitchings of the HARPS CCD, as for the case of HD88725. We noticed this when fitting and removing from the RVs a sinusoid with a period of 1\,year decreased the power of the 245\,d peak.

\begin{figure}
  \centering
  \includegraphics[width=\hsize]{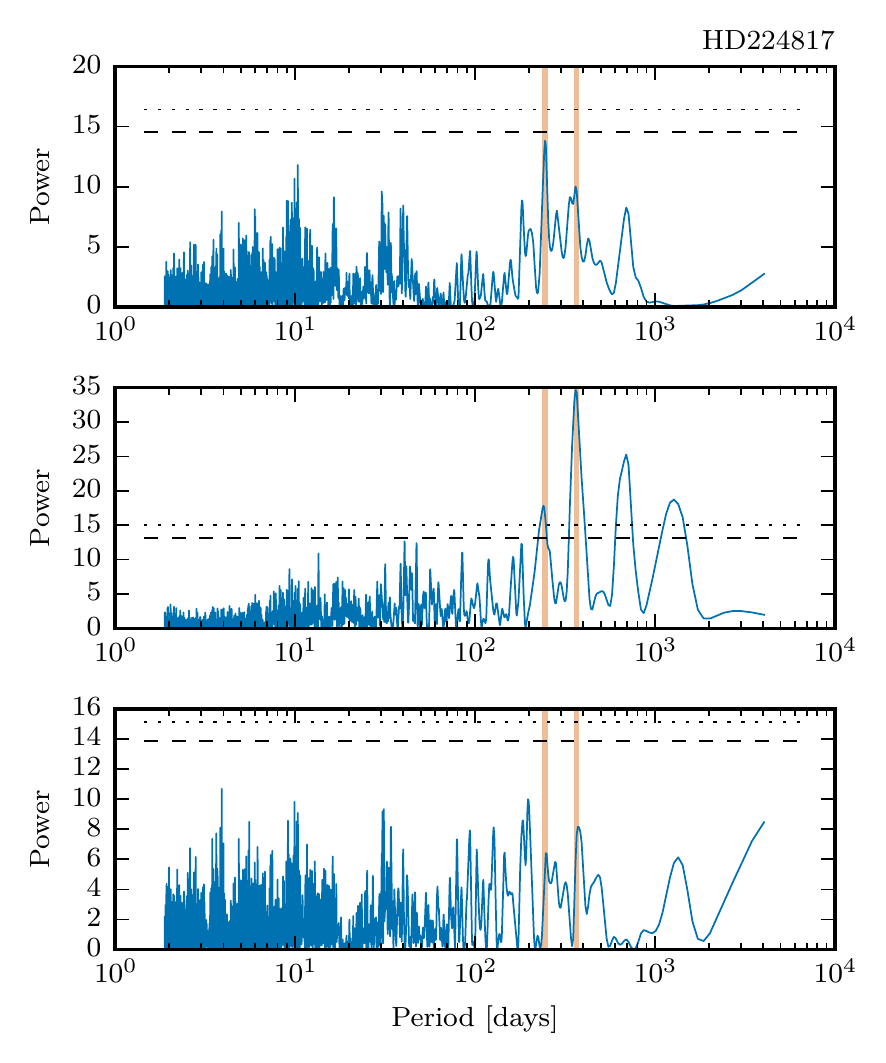}
    \caption{Periodograms of the original radial velocities of HD224817 (top panel), the RVs derived with a correlation mask that contains only the spectral lines crossing the CCD block stitchings (middle panel), and those derived using a correlation mask without those lines (bottom panel). The vertical lines show the period of the highest peak in the top periodogram and the orbital period of the Earth.}
       \label{fig:HD224817-gaps}
  \end{figure}

  To further check that the signal is indeed caused by the instrument, we created different correlation masks both with and without the lines falling close to the block stitchings \citep{Dumusque2015}. 
  The periodograms for the original RVs and those from the two custom masks are shown in Fig. \ref{fig:HD224817-gaps}.
  When the spectral lines that cross the block stitchings are removed from the correlation mask, the 245\,d peak vanishes from the periodogram (bottom panel).
  The periodogram of the corrected RVs does not show significant peaks and we do not find evidence for the presence of Keplerian signals.

  The second highest peak in the periodogram of the original radial velocities, around 10\,d, is not far from the estimated rotation period of the star (Table \ref{table:parameters}). A clear (though not significant) peak at this period is present in the periodogram of the FWHM indicator.
  We conclude that the data available for HD224817 does not suggest the presence of any planetary companions.

\subsection{Stars without identified signals}\label{sec:no-peaks}

\paragraph{HD31128} 
  This star was followed very closely with a total of 215 radial-velocity measurements on 169 different nights. 
  These numbers already take into account an outlier observation at $\text{BJD}=2455304.48$ which was removed prior to the analysis. 
  This observation has a very low CCF contrast value, making us suspect of some instrumental or observational error at this date.

  The RV data show a high dispersion though the average uncertainty is also above $3\ms$. 
  These can be explained by the high effective temperature and very low metallicity of this star, which hinder the precise calculation of the radial-velocity with the CCF technique (low metallicity stars within this temperature range have shallower lines, making the determination of precise RVs more difficult, while the higher temperature introduces higher oscillation and granulation noise levels; \citealt{Dumusque2011a})

  Despite the large number of measurements, none of the periodogram peaks are significant (Fig. \ref{fig:HD31128}). 
  According to the model selection analysis, there is no evidence for planetary companions orbiting this star.
  The constant model is preferred relative to the 1-planet model according to the value of the evidence (with an odds ratio of 2.5) and the BIC (with $\Delta$BIC of 5.4). However, the AIC$_C$ selects instead the 2-planet model but, in light of our general results (see Sect. \ref{sec:comparison_model_selection}), we tend to trust the Bayesian criteria.

\paragraph{HD119173} 
  This star was observed on 98 different nights (totalling 128 measurements) which cover a total timespan of 8.1 years. 
  No significant periodogram peaks were found (Fig. \ref{fig:HD119173}), and in the model selection analysis the constant model is preferred over the 1-planet model (with odds ratio $\mathcal{O}_{01} \sim 4.5$ and $\Delta$BIC $=8.7$). We conclude that there is no evidence for the presence of planetary signals in these data.

\paragraph{HD119949} 
  The periodogram of the 81 measurements, shown in Fig. \ref{fig:HD119949}, does not present any significant peaks. 
  The Bayesian search for Keplerian signals reaches the same conclusions (odds ratio $\mathcal{O}_{01} \sim 4$). The AIC$_C$ again selects a more complex 1-planet model with a non-significant $\Delta$AIC$_C$. We report no clear evidence for planetary companions.

\paragraph{HD126793} 
  A total of 96 measurements on 79 different nights were obtained. 
  The periodogram (Fig. \ref{fig:HD126793}) does not show any significant peaks and, in this case as well, the model comparison does not find evidence for Keplerian signals in the data. For these data, all model selection criteria select the constant model as the most probable with $\mathcal{O}_{01} \sim 4$, $\Delta$AIC$_C$ $=1.1$, $\Delta$BIC $=11.5$.

\section{Comparison of model selection criteria}\label{sec:comparison_model_selection}

For the maximum-likelihood solutions, the AIC$_C$ and BIC can be calculated using the number of free parameters in a given model. These are sometimes used as model selection criteria, and are much easier to calculate than the value of the evidence. In Table \ref{table:model-selection} we show the AIC$_C$ and BIC values for each star and model, using the maximum-likelihood obtained from all {\sc MultiNest} samples. We highlight in bold the best model according to each criteria.

As general trends in our results we can note that the AIC$_C$ tends to choose a model with a higher number of Keplerian signals.
The BIC is found to be more conservative and tends to agree with the model with the highest evidence. These results suggest that the BIC can be used as a viable approximation to the evidence, given that the (global) maximum of the likelihood function can be found.


\begin{table*}
\caption{Model selection results for the stars without significant periodogram peaks. 
         For each star and each model we show the logarithm of the evidence, the odds ratio with respect to the constant model, 
         the (normalised) value of the posterior distribution $\p[n \given d]$ and the AIC$_C$ and BIC for the maximum-likelihood solution.}             
\label{table:model-selection}      
\centering          
\begin{tabular}{llcclcc}
        &               & $\ln E$         & $\mathcal{O}_{i0}$ & \multicolumn{1}{c}{$\p[n \given d]$\tablefootmark{1}}      & AIC$_C$\tablefootmark{\,2} & BIC\tablefootmark{\,2} \\
HD21132  &   &   &   &   &   &   \\
         & constant     & -229.5          & 1           & \rule{0.194cm}{1.5ex}        & 452.1           & \textbf{456.9} \\
         & 1 keplerian  & \textbf{-227.7} & 6.13        & \rule{1.000cm}{1.5ex}        & 443.7           & 459.3          \\ 
         & 2 keplerians & -229.0          & 1.54        & \rule{0.472cm}{1.5ex}        & \textbf{438.5}  & 463.5          \\[1em]
HD22879  &              &                 &               &                              &                   &                 \\
         & constant     & -164.8          & 1             & \rule{0.067cm}{1.5ex}        & 322.2            & 326.8           \\
         & 1 keplerian  & \textbf{-162.8} & 7.87          & \rule{1.000cm}{1.5ex}        & \textbf{300.8}   & \textbf{315.9}  \\
         & 2 keplerians & -164.3          & 1.76          & \rule{0.395cm}{1.5ex}        & 303.3            & 327.2           \\[1em]       
HD31128  &   &   &   &   &   &   \\
         & constant     & \textbf{-473.9} & 1             & \rule{1.000cm}{1.5ex}        & 940.7           & \textbf{946.9}   \\
         & 1 keplerian  & -474.9          & 0.40          & \rule{0.665cm}{1.5ex}        & 931.0           & 952.3   \\
         & 2 keplerians & -475.7          & 0.17          & \rule{0.245cm}{1.5ex}        & \textbf{919.1}  & 954.8   \\[1em]
HD87838  &              &                 &               &                              &     &     \\
         & constant     & -214.5          & 1             & \rule{0.474cm}{1.5ex}        & 421.4           & \textbf{426.2}  \\
         & 1 keplerian  & \textbf{-213.9} & 1.75          & \rule{1.000cm}{1.5ex}        & \textbf{410.8}  & 426.6           \\
         & 2 keplerians & -215.1          & 0.56          & \rule{0.582cm}{1.5ex}        & 411.2           & 436.6           \\[1em]
HD119173  &   &   &   &   &   &   \\
         & constant     & \textbf{-202.0} & 1             & \rule{1.000cm}{1.5ex}        & 397.1            & \textbf{402.1}    \\
         & 1 keplerian  & -203.5          & 0.23          & \rule{0.115cm}{1.5ex}        & \textbf{393.9}   & 410.8    \\
         & 2 keplerians & -205.8          & 0.02          & \rule{0.031cm}{1.5ex}        & 402.6            & 429.9    \\[1em]
HD119949  &   &   &   &   &   &   \\
         & constant     & \textbf{-188.5} & 1             & \rule{1.000cm}{1.5ex}        & 370.1           & \textbf{374.8}  \\
         & 1 keplerian  & -189.9          & 0.25          & \rule{0.178cm}{1.5ex}        & \textbf{366.8}  & 382.0  \\
         & 2 keplerians & -191.8          & 0.03          & \rule{0.058cm}{1.5ex}        & 377.6           & 401.7  \\[1em]
HD126793  &   &   &   &   &   &   \\
         & constant     & \textbf{-174.1} & 1             & \rule{1.000cm}{1.5ex}        & \textbf{341.0} & \textbf{345.6}    \\
         & 1 keplerian  & -175.6          & 0.23          & \rule{0.225cm}{1.5ex}        & 342.1          & 357.1    \\
         & 2 keplerians & -176.8          & 0.07          & \rule{0.058cm}{1.5ex}        & 350.3          & 373.9    \\[1em]
HD224817 &              &                 &               &                              &                 &                 \\
         & constant     & -218.1          & 1             & \rule{0.327cm}{1.5ex}        & 428.8           & \textbf{433.9}  \\
         & 1 keplerian  & \textbf{-217.8} & 1.25          & \rule{1.000cm}{1.5ex}        & 418.3           & 435.3           \\
         & 2 keplerians & -218.9          & 0.40          & \rule{0.593cm}{1.5ex}        & \textbf{416.2}  & 443.9           \\[1em]

\end{tabular}
\tablefoot{
\tablefoottext{1}{The bars represent the posterior distribution for the number of keplerians in the model. For each star, this posterior is normalized to the value in the most probable bin.}
\tablefoottext{2}{Calculated using maximum likelihood value from all MultiNest samples and the number of free parameters for each model.}
}
\end{table*}


\section{Detection Limits}\label{sec:detection-limits}

Many of the stars we analysed in Sect \ref{sec:results} do not contain significant periodic signals or show signals originating from activity and sampling contaminations. 
In this section, we derive upper limits to the planetary signals that can be present and yet still undetectable in these data. For a given star we determine which planet, as a function of its mass and orbital period, can already be ruled out with our observations.

The detection limits are calculated with a standard procedure: injecting trial circular orbits into the observed data \citep{Cumming1999,Endl2001,Zechmeister2009,Dumusque2011b,Mayor2011,Mortier2012}. We explore all periods in the range from 1\,d to 500\,d and semi-amplitudes from 0 to 10$\ms$ (using a binary search). For 10 linearly-spaced phases, we compare the periodogram power of the injected period with the FAP level of that period in the original dataset. If, for all phases, the former is higher we consider the planet to be detected. We convert the semi-amplitude to planetary mass using the stellar masses listed in Table \ref{table:parameters}. 

It is important to note that this method assumes the original dataset (in which mock planets are injected) to contain only non-correlated noise. For the stars analysed in Sect. \ref{sec:no-peaks}, this assumption is valid. For the stars analysed in Sect. \ref{sec:unconfirmed-signals}, we substracted the putative orbital solution from the data before calculating the detection limits. Since correcting for activity-induced signals and artifacts originating from the sampling is much more prone to error, for the stars discussed in Sects. \ref{sec:signals-activity} and \ref{sec:other-cases}, we assume that the observed radial-velocities contain only noise. Therefore, for this last group of stars, the detection limits estimates can be taken as conservative.

\begin{figure*}
\centering
\includegraphics[width=\hsize]{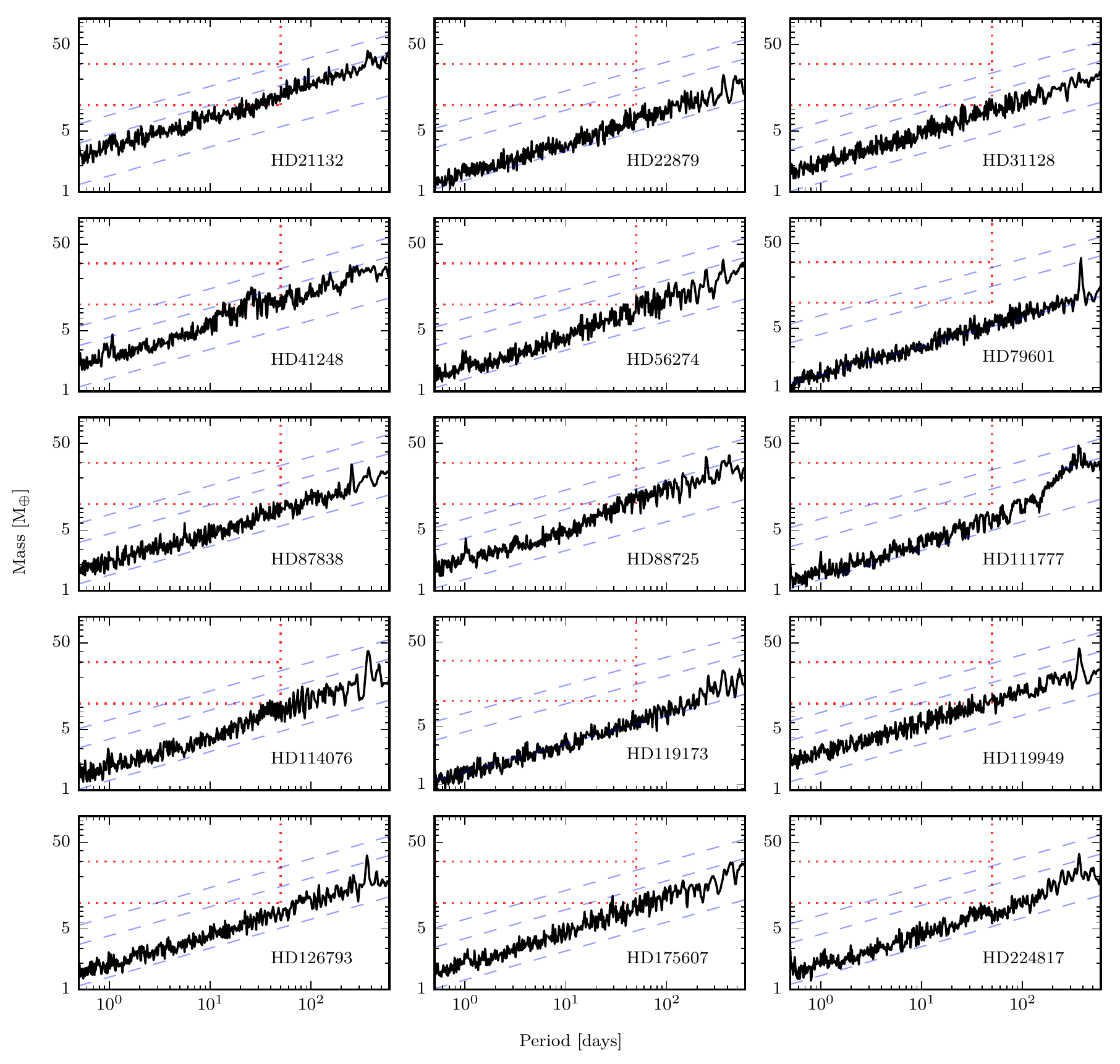}
  \caption{Detection limits for the 15 stars studied in this work. The plots show planetary mass against orbital period. The thik black line shows the 99\% detection limits, estimated by injecting trial circular orbits. The dashed blue lines represent circular planetary signals with RV semi-amplitudes of 1, 3 and 5 $\ms$ (from bottom to top in each panel). The regions delimited by the red dotted lines correspond to periods lower than $50\,$d, and masses in the ranges $M=10-30\earthmass$ and $M=30-100\earthmass$.
}
     \label{fig:detection-limits-all}
\end{figure*}

The detection limits for all 15 stars are shown in Fig. \ref{fig:detection-limits-all}. The plots show minimum mass, in Earth masses, against orbital period. The black curves correspond to the 99\% detection limits and the blue dashed curves indicate circular planetary signals with RV semi-amplitudes of 1, 3 and 5 $\ms$.
The red dotted regions correspond to planets with $P<50\,$d, and masses in the ranges $M=10-30\earthmass$ and $M=30-100\earthmass$. These can be compared to the same regions considered in \citep{Mayor2011} for the calculation of the planet ocurrence rate. 

Almost uniformly in this sample, we are sensitive to planets more massive than $10\earthmass$ up to orbital periods of 50 days.
For shorter period planets, with $P\sim10$ days and smaller, the current data would already allow for the detection of planets with just a few Earth masses.

\section{Discussion and conclusions}\label{sec:discussion}

From over 100 targets in our metal-poor large program, we selected those which were observed on more than 75 nights. A homogeneous analysis of the radial-velocity observations was carried out by comparing models with a varying number of Keplerian signals. We used the nested sampling algorithm to estimate the evidence of a given model and compared the results with the posterior distribution of the number of Keplerians $n$. We also calculate the AIC$_C$ and BIC and analyse the periodograms of the radial velocities and activity indicators.

We find, in some cases, a disagreement between the evidence results and the presence of significant periodogram peaks. 
The Bayesian analysis was able to identify a few Keplerian signals which are not significant in the periodograms. 
Nevertheless, the values of the odds ratios indicate that these detections are not very significant. 
The results obtained with {\sc MultiNest} almost always\footnote{The exceptions are the 2-planet models for HD87838 and HD224817. 
This can be due to poorer sampling by one of the algorithms or incomplete convergence.} agree with the posterior distribution $\p[n \given d]$, which is reassuring but not surprising given that the same priors and likelihood were used in both algorithms.

Signals induced by stellar activity were detected for HD21132, HD56274 and HD114076, besides the known case of HD41248. After attempting to correct for these signals, we were not able to recover any additional signs of planetary companions.

For HD79601, HD22879 and HD111777, the time sampling of the observations induces spurious signals in the radial-velocities, hindering the detection of planetary signals. 
We also identified a clear 1 year periodicity and its harmonics, caused by instrumental effects, on the radial velocities of HD88725 and HD224817.
These instrument-induced RV variations are not expected to be present in all stars since each star may have different values of the BERV and different spectral lines crossing the CCD block stitchings \citep[see][]{Dumusque2015}.

HD87838 shows an interesting signal around 68\,d, which can be fitted with a slightly eccentric Keplerian.
The model selection does not provide strong evidence for the presence of this planet but we hope to obtain more data to assert its nature.

It is important to note that the value of the evidence (and our estimate) is sensitive to the priors on the parameters of a given model. 
This is not necessarily a limitation, meaning only that care should be taken when choosing the priors and that they should be stated clearly in any analysis. 

As an example of these points, we re-analysed the data of HD87838 with $n=1$, considering the prior for the orbital period to be uniform between 65\,d and 75\,d
\footnote{The example is contrived but might shed light on why one should not adjust the priors after having looked at the data (or the periodogram, for that matter).}. The resulting value for the evidence was found to be $-209.5$, in contrast to the value determined previously, of $-213.9$ (Table \ref{table:model-selection}). This corresponds to an odds ratio of $\sim\!\!156$ against the constant model, which would imply ``very strong'' evidence for a Keplerian signal and a confident planet detection.

We performed a sensitivity analysis of our results, namely by considering different priors for the orbital periods and semi-amplitudes. Even if, in some cases, a different model would be preferred (that is, yield a higher evidence), the absolute value of the Bayes factors did not change considerably. This means that our level of confidence in the presence or not of a planetary signal does not depend severely on the parameter priors.

Even though our working sample is small, it already allows for an analysis of the frequency of planets around these stars. We do not attempt to provide constraints on the planet frequency as a function of stellar metallicity, since our sample does not span completely the metallicity space (Fig. \ref{fig:sample-distribution}). 

The probability of obtaining $k$ detections in a sample of size $N$ is given by the binomial distribution, assuming a planet frequency $f_p$
\begin{equation}
	p(k\,|\,N, f_p) = \frac{N!}{k!(N-k)!} \,f_p^{\,k} \,(1-f_p)^{N-k}
\end{equation}
This equation can be thought of as a function of $f_p$ for given values of $k$ and $N$, therefore representing a (un-normalised) posterior distribution for $f_p$. The number of planets detected in a given sample places a constraint on the probable values of the planet frequency, which can be expressed by the mode and the range that covers, e.g., 68\% of the distribution. 

Considering the detection of the planet orbiting HD175607, in a sample of 15 stars, this procedure gives a constraint of 
$f_p = 6.67\%^{+4.16\%}_{-5.58\%}$.
Taking into account also the unconfirmed planet around HD87838, one obtains 
$f_p = 13.33\%^{+5.80\%}_{-8.84\%}$. 
This last result is in agreement with that of \citet{Mayor2011}, who found $12.27\pm2.45\%$ for the combined occurrence rate of planets with orbital periods smaller than 50 days and masses in the range $M=10-100\earthmass$. 
We nevertheless stress that our constraints on the planet occurrence rate are preliminary, since they are based on a small sample and subject to selection biases.

Recent studies have identified correlations between the properties of the planetary system and the metallicity of the star. Lower metallicity stars have been found to host longer period planets \citep{Beauge2013,Adibekyan2013} or lower eccentricity orbits \citep{Dawson2013}. These results, together with the detection limits calculated in Sect \ref{sec:detection-limits}, might explain why few planets have been detected so far in our sample. The search for low-mass planets around metal-poor stars requires not only high-precision radial velocities but also a very long baseline of observations.

Our results show that the detection of small planets in these data is hindered by time sampling issues and activity-induced signals. 
Careful planning of the observational strategy and detailed analyses of the aliasing structures are necessary to mitigate some of these problems and we must seek ways to make use of all the information contained in the RVs and in all the activity indicators, in an optimal and physically-meaningful way.

The analysis of the full sample of 109 metal-poor stars is ongoing and we expect to provide constraints on the frequency of planets orbiting these stars when the observations of the Large Program are finished.

\begin{acknowledgements}
This work was supported by Funda\c{c}\~ao para a Ci\^encia e a Tecnologia (FCT) through the research grant UID/FIS/04434/2013.
We also acknowledge the support from FCT in the form of grant reference PTDC/FIS-AST/1526/2014.
JPF acknowledges the support from FCT through the grant reference SFRH/BD/93848/2013 and thanks Brendon Brewer for his assistance with the diffusive nested sampling code.
PF and NCS acknowledge support by FCT through Investigador FCT contracts of reference IF/01037/2013 and IF/00169/2012, respectively, and POPH/FSE (EC) by FEDER funding through the program ``Programa Operacional de Factores de Competitividade - COMPETE''. PF further acknowledges support from FCT in the form of an exploratory project of reference IF/01037/2013CP1191/CT0001.
AM received funding from the European Union Seventh Framework Program (FP7/2007-2013) under grant agreement number 313014 (ETAEARTH).
AS is supported by the European Union under a Marie Curie Intra-European Fellowship for Career Development with reference FP7-PEOPLE-2013-IEF, number 627202.

\end{acknowledgements}

\bibliographystyle{aa}
\bibliography{/home/joao/phd/bib/zotero_library_AAabbrev}{}

\Online
\begin{appendix} 
\section{RV time series and periodograms}

This appendix contains figures with the radial-velocity time series and respective GLS periodogram for each star.


	\begin{figure}[h]
	\includegraphics[width=0.9\hsize]{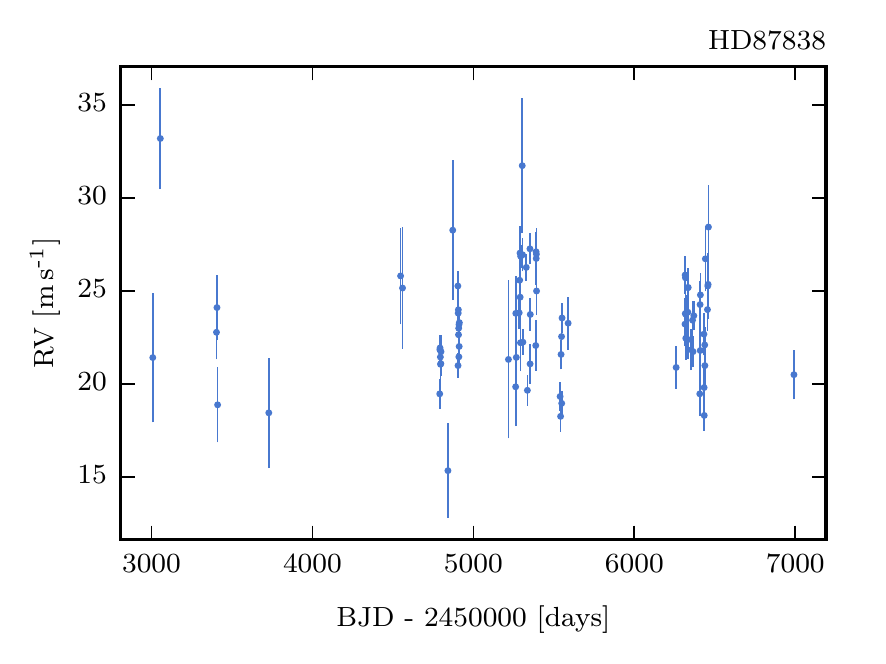}
	\includegraphics[width=0.9\hsize]{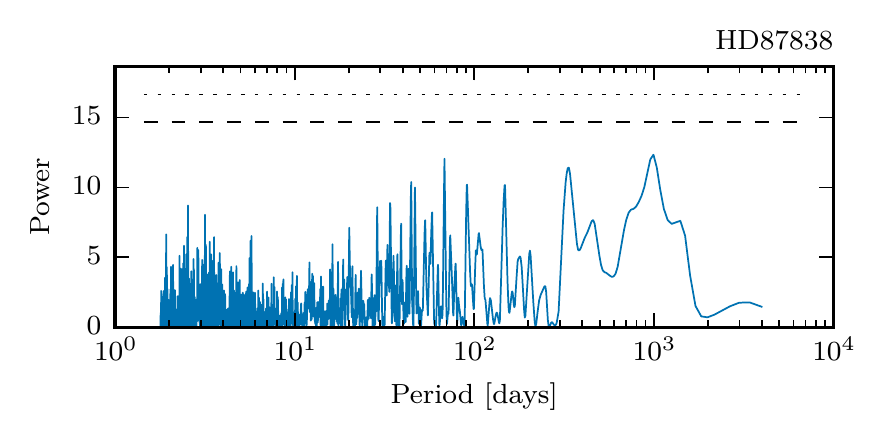}
	\caption{Radial-velocity time series and periodogram for HD87838.}
	\label{fig:HD87838}
	\end{figure}

	\begin{figure}[h]
	\includegraphics[width=0.9\hsize]{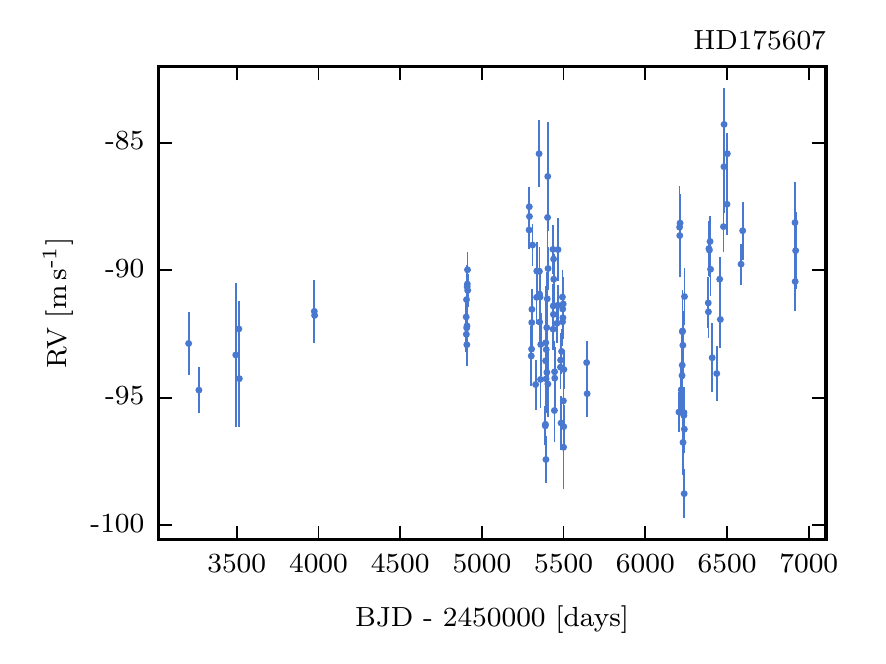}
	\includegraphics[width=0.9\hsize]{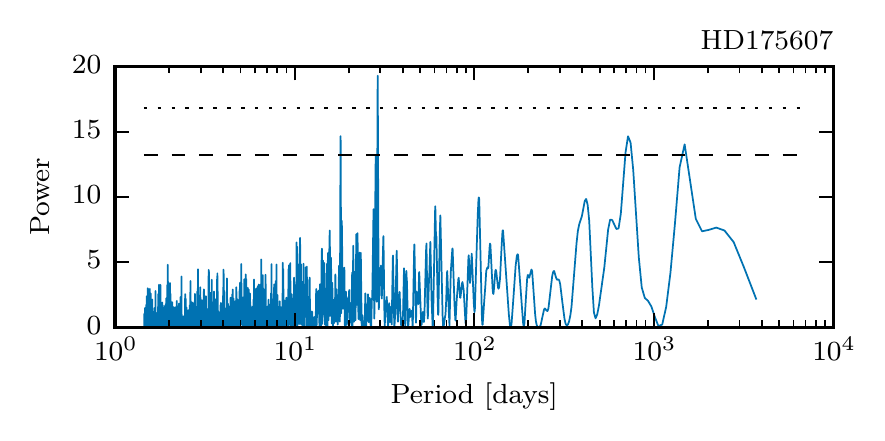}
	\caption{Radial-velocity time series and periodogram for HD175607.}
	\label{fig:HD175607}
	\end{figure}

	\begin{figure}[h]
	\includegraphics[width=0.9\hsize]{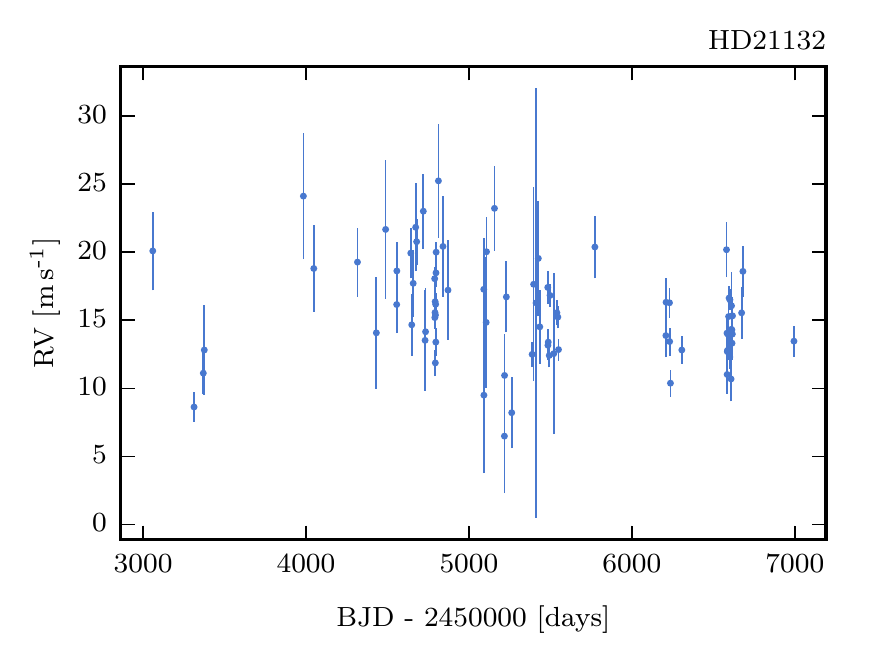}
	\includegraphics[width=0.9\hsize]{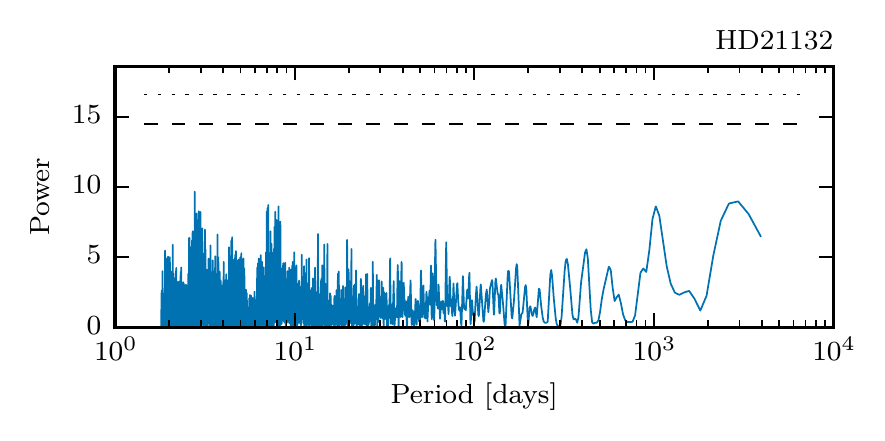}
	\caption{Radial-velocity time series and periodogram for HD21132.}
	\label{fig:HD21132}
	\end{figure}

	\begin{figure}[h]
	\includegraphics[width=0.9\hsize]{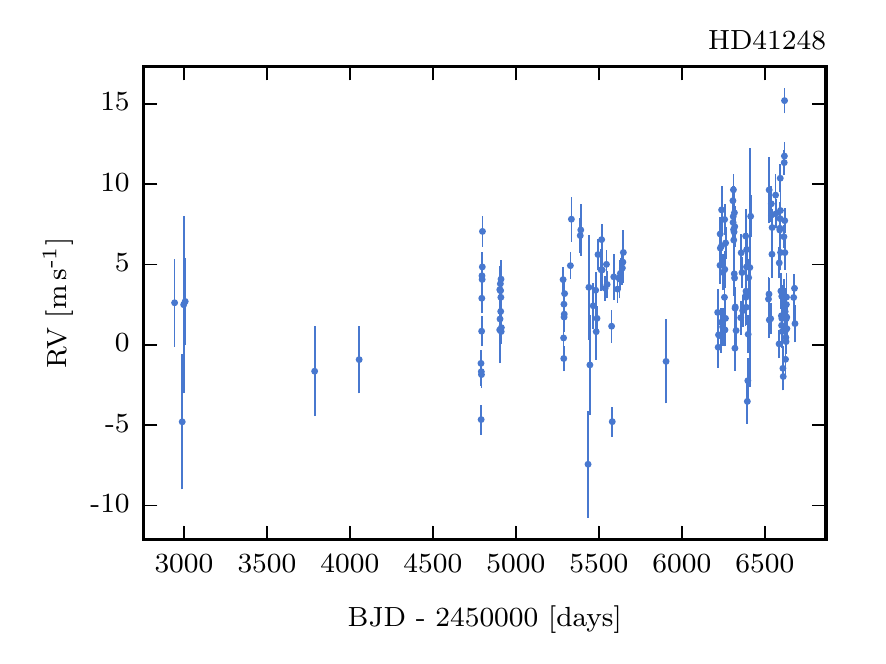}
	\includegraphics[width=0.9\hsize]{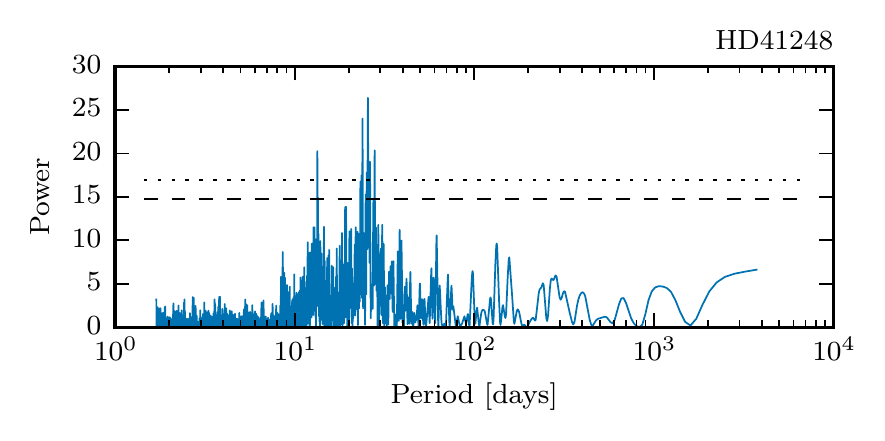}
	\caption{Radial-velocity time series and periodogram for HD41248.}
	\label{fig:HD41248}
	\end{figure}

	\begin{figure}[h]
	\includegraphics[width=0.9\hsize]{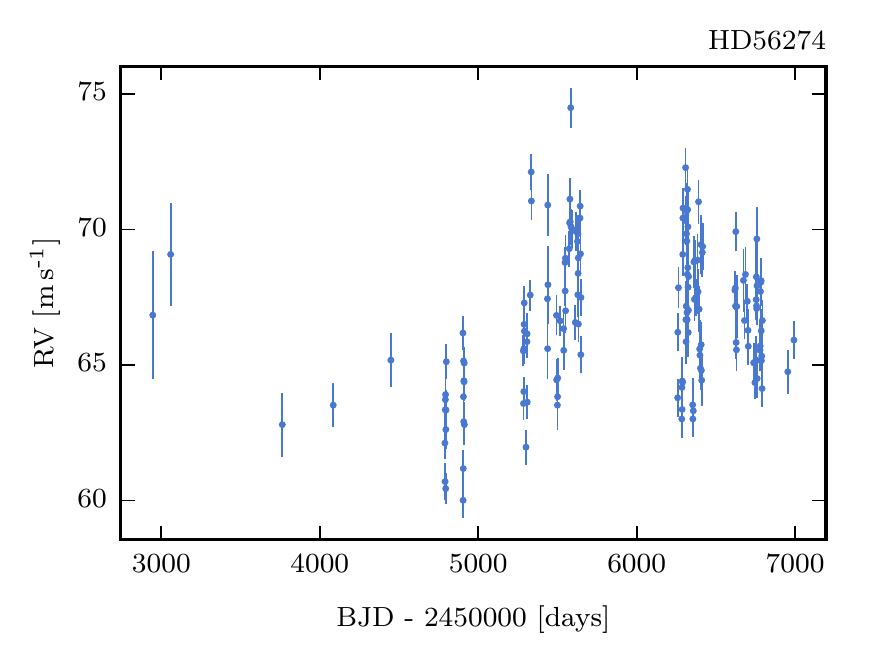}
	\includegraphics[width=0.9\hsize]{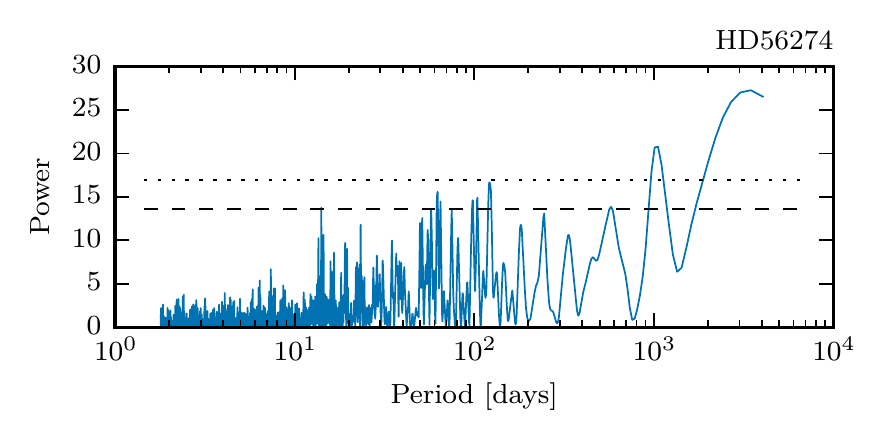}
	\caption{Radial-velocity time series and periodogram for HD56274.}
	\label{fig:HD56274}
	\end{figure}

	\begin{figure}[h]
	\includegraphics[width=0.9\hsize]{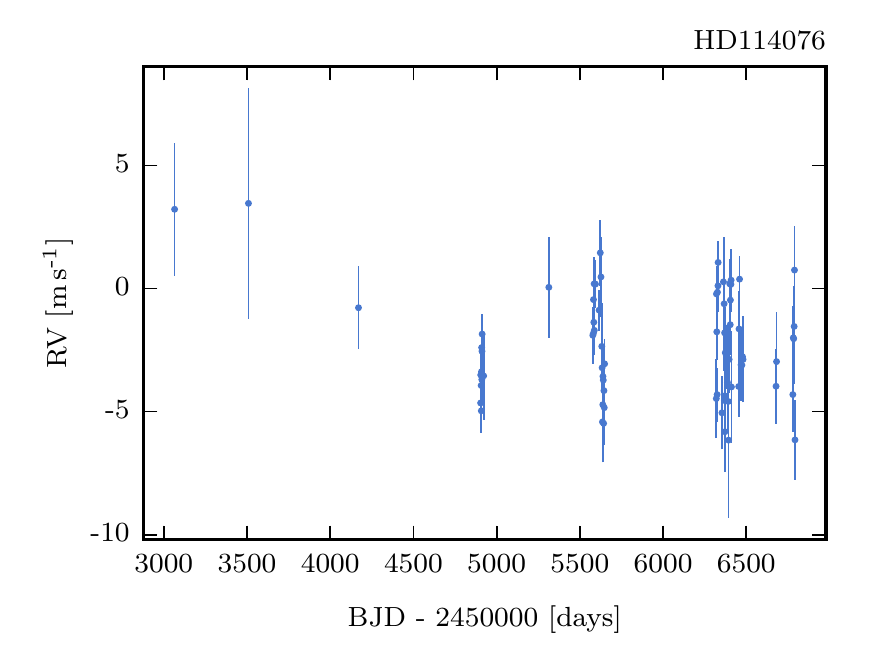}
	\includegraphics[width=0.9\hsize]{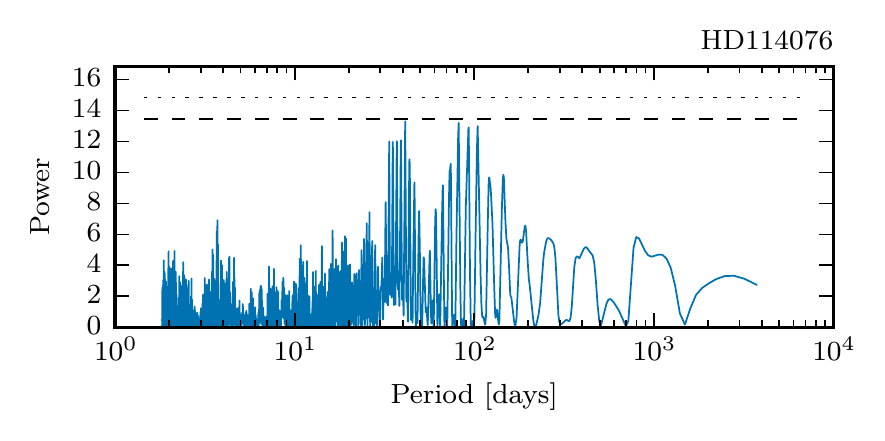}
	\caption{Radial-velocity time series and periodogram for HD114076.}
	\label{fig:HD114076}
	\end{figure}

	\begin{figure}[h]
	\includegraphics[width=0.9\hsize]{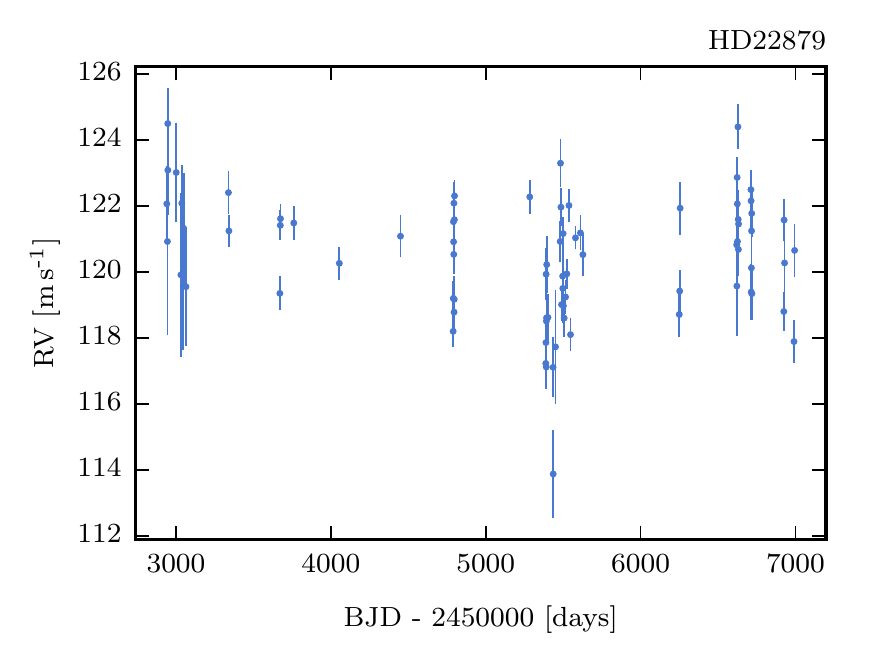}
	\includegraphics[width=0.9\hsize]{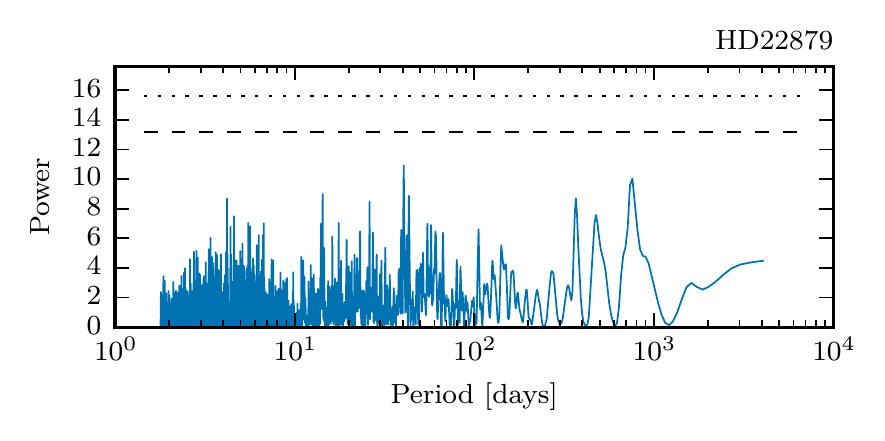}
	\caption{Radial-velocity time series and periodogram for HD22879.}
	\label{fig:HD22879}
	\end{figure}

	\begin{figure}[h]
	\includegraphics[width=0.9\hsize]{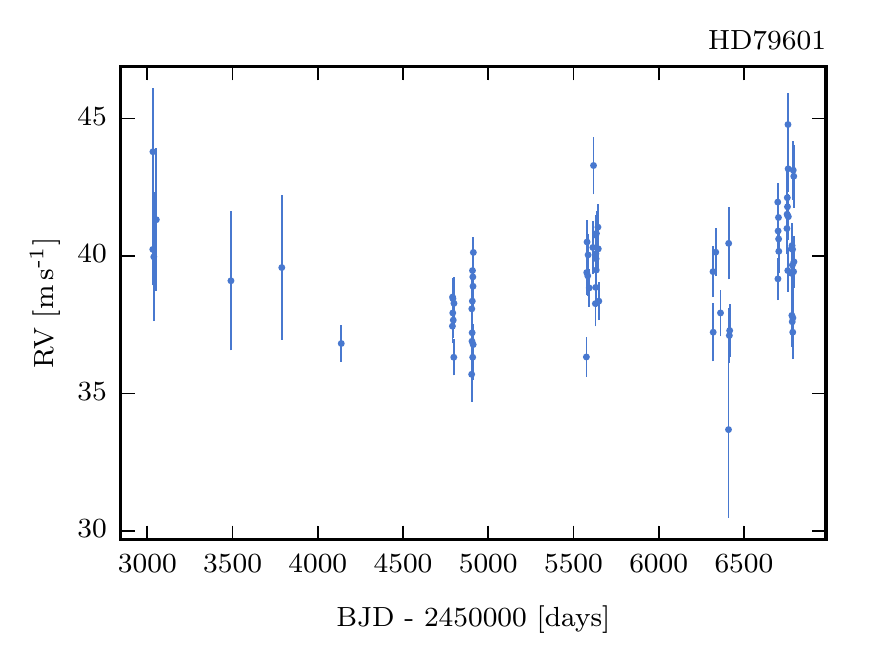}
	\includegraphics[width=0.9\hsize]{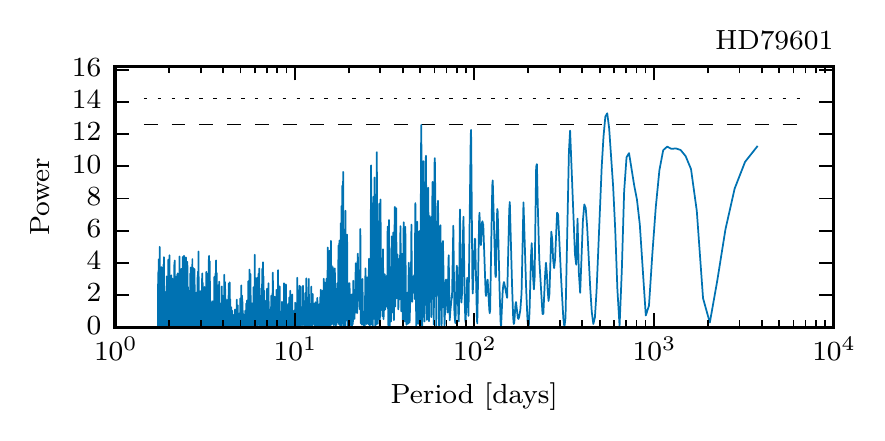}
	\caption{Radial-velocity time series and periodogram for HD79601.}
	\label{fig:HD79601}
	\end{figure}

	\begin{figure}[h]
	\includegraphics[width=0.9\hsize]{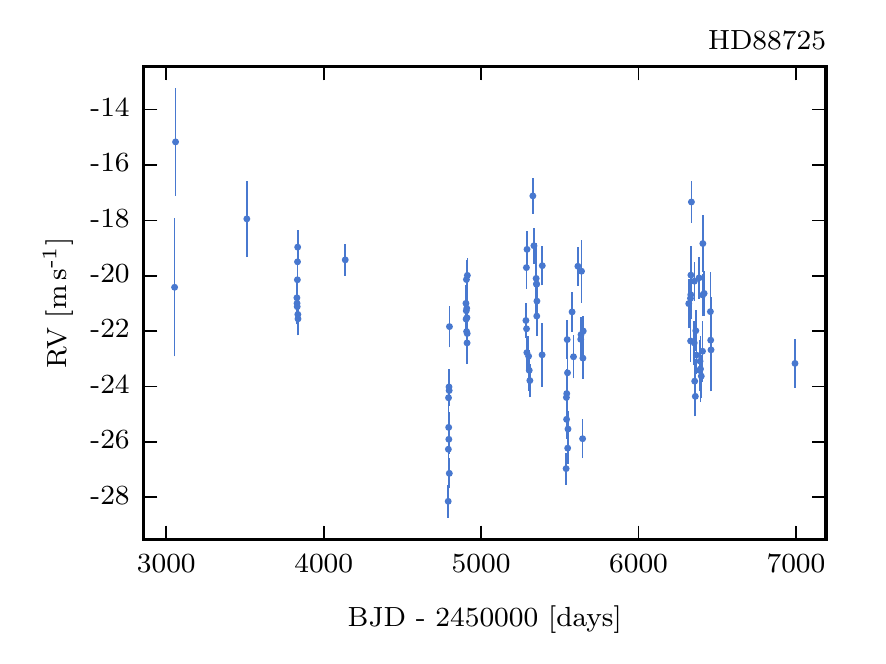}
	\includegraphics[width=0.9\hsize]{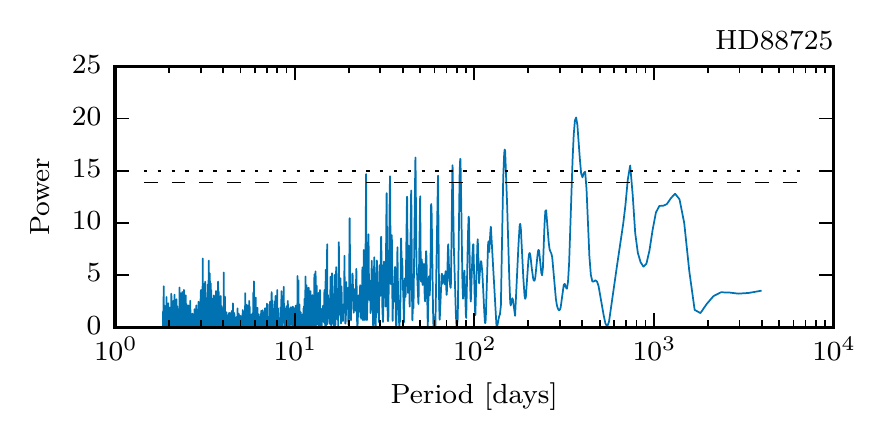}
	\caption{Radial-velocity time series and periodogram for HD88725.}
	\label{fig:HD88725}
	\end{figure}

	\begin{figure}[h]
	\includegraphics[width=0.9\hsize]{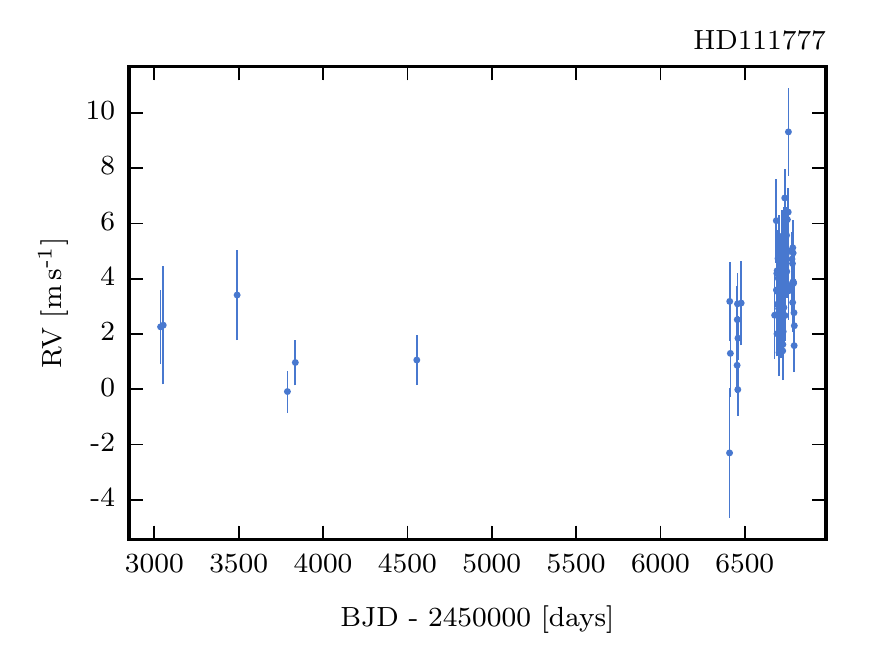}
	\includegraphics[width=0.9\hsize]{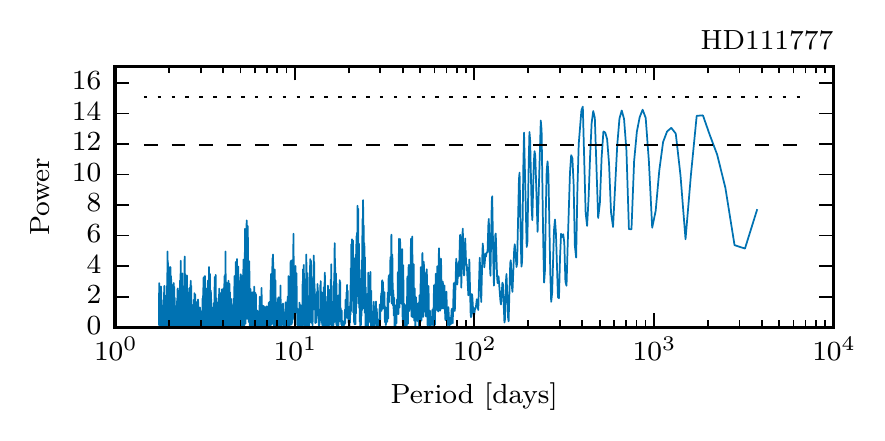}
	\caption{Radial-velocity time series and periodogram for HD111777.}
	\label{fig:HD111777}
	\end{figure}

	\begin{figure}[h]
	\includegraphics[width=0.9\hsize]{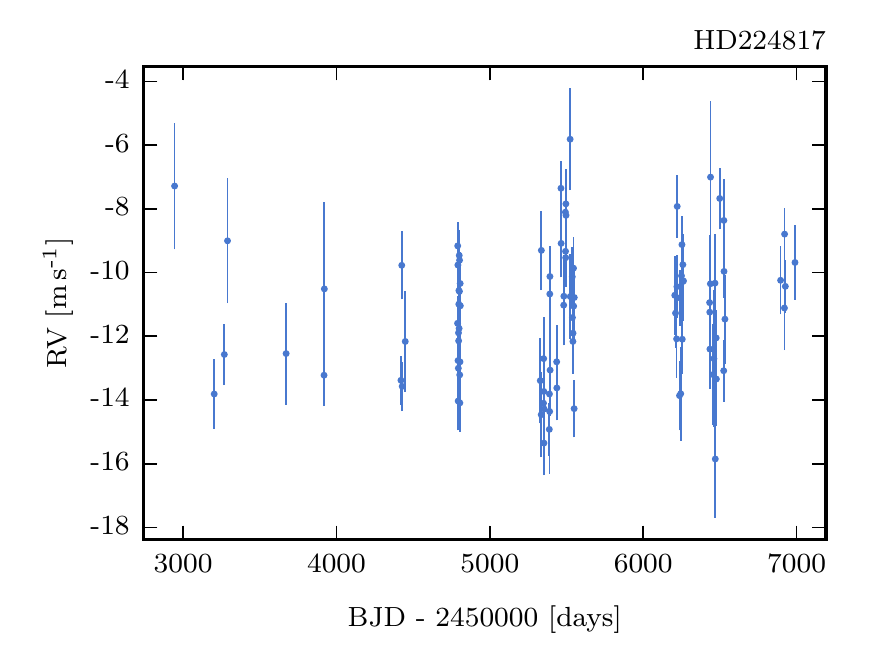}
	\includegraphics[width=0.9\hsize]{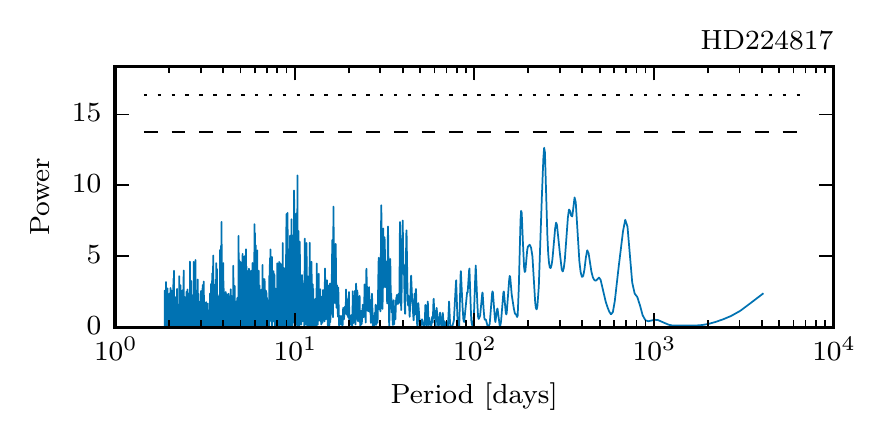}
	\caption{Radial-velocity time series and periodogram for HD224817.}
	\label{fig:HD224817}
	\end{figure}

	\begin{figure}[h]
	\includegraphics[width=0.9\hsize]{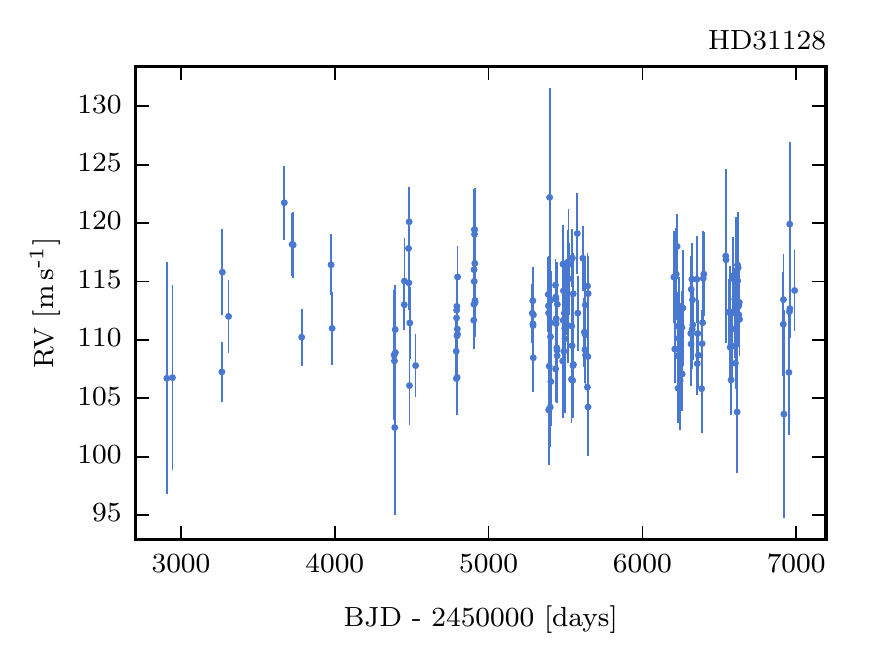}
	\includegraphics[width=0.9\hsize]{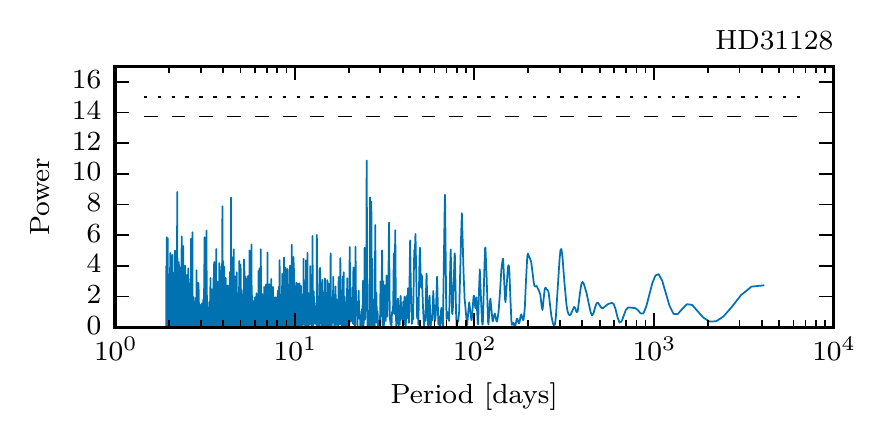}
	\caption{Radial-velocity time series and periodogram for HD31128.}
	\label{fig:HD31128}
	\end{figure}

	\begin{figure}[h]
	\includegraphics[width=0.9\hsize]{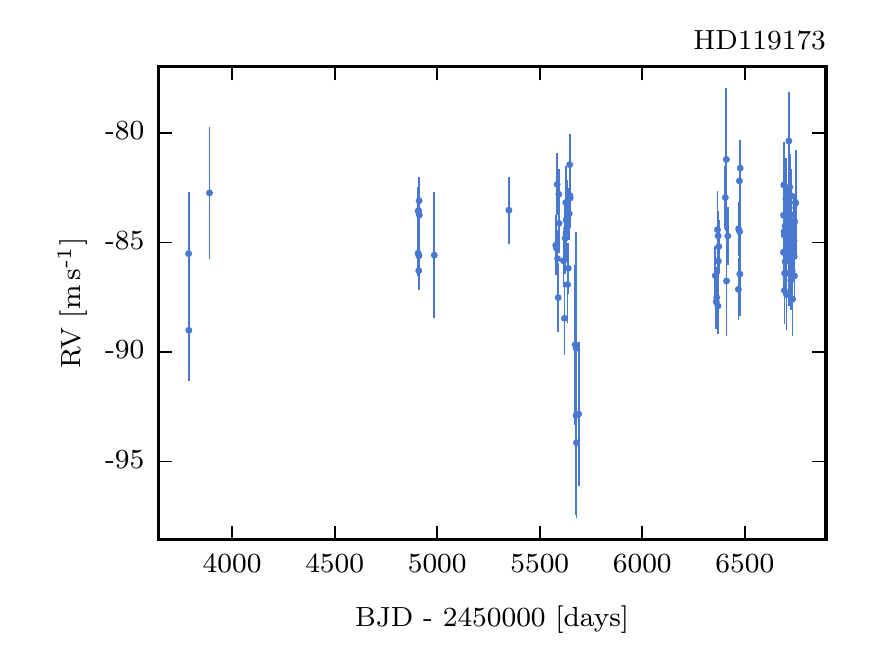}
	\includegraphics[width=0.9\hsize]{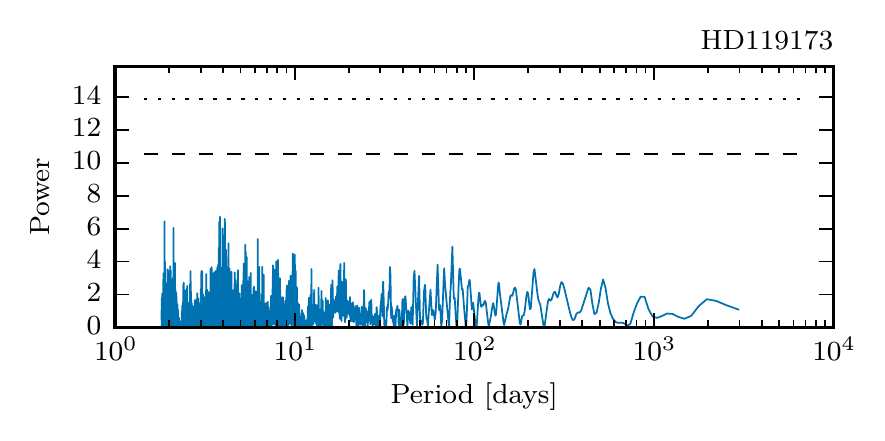}
	\caption{Radial-velocity time series and periodogram for HD119173.}
	\label{fig:HD119173}
	\end{figure}

	\begin{figure}[h]
	\includegraphics[width=0.9\hsize]{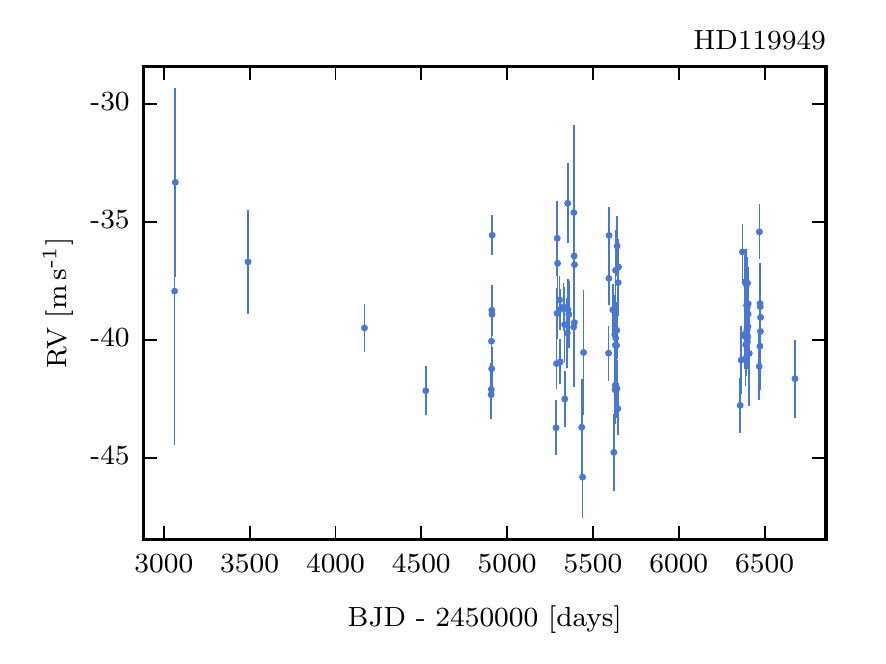}
	\includegraphics[width=0.9\hsize]{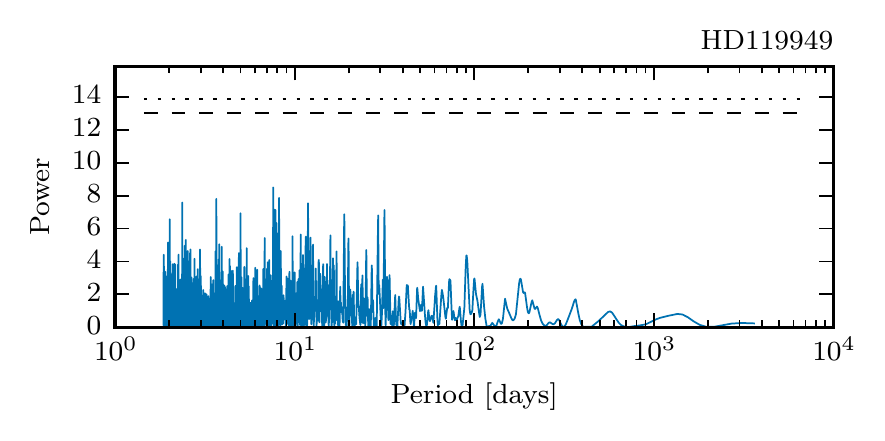}
	\caption{Radial-velocity time series and periodogram for HD119949.}
	\label{fig:HD119949}
	\end{figure}

	\begin{figure}[h]
	\includegraphics[width=0.9\hsize]{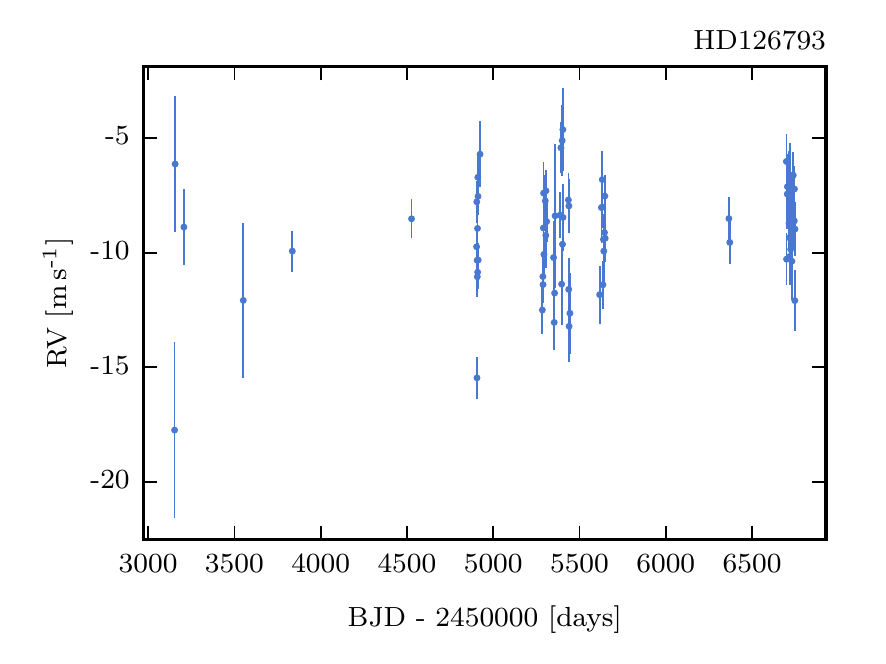}
	\includegraphics[width=0.9\hsize]{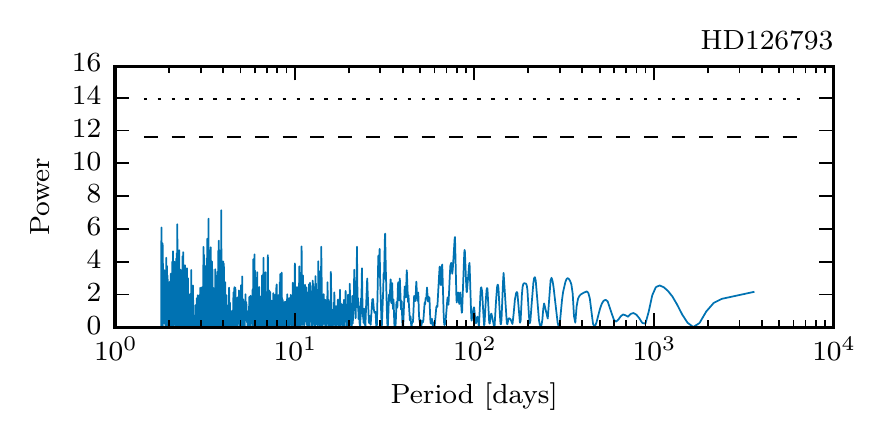}
	\caption{Radial-velocity time series and periodogram for HD126793.}
	\label{fig:HD126793}
	\end{figure}

\end{appendix}

\end{document}